\documentclass[showpacs,preprintnumbers,amsmath,amssymb,prb]{revtex4}

\usepackage{graphicx}
\usepackage{dcolumn}
\usepackage{bm}
\usepackage{amsfonts}
\usepackage{amssymb}

\newcommand{\no}[1]{}

\def\drawline#1#2{\raise 2.5pt\vbox{\hrule width #1pt height #2pt}}

\def\trian{\raise 1.25pt\hbox{$\scriptscriptstyle\triangle$}\nobreak\ }

\def\square{${\vcenter{\hrule height .4pt
        \hbox{\vrule width .4pt height 3pt \kern 3pt
        \vrule width .4pt}
        \hrule height .4pt}}$\nobreak\ }

\def\plus{\raise 1.25pt \hbox{$\scriptscriptstyle +$}\nobreak\ }

\begin{document}

\title{Cloud microphysical effects of turbulent mixing and entrainment}
\author{Bipin Kumar$^1$, J\"org Schumacher$^1$ and Raymond A. Shaw$^2$}
\affiliation{$^1$ Institute of Thermodynamics and Fluid Mechanics, Ilmenau University of Technology, D-98684 Ilmenau, Germany \\
$^2$ Department of Physics, Michigan Technological University, Houghton, MI 49931, USA}

\date{\today}

\begin{abstract}
Turbulent mixing and entrainment at the boundary of a cloud is studied by means of direct numerical simulations that couple the Eulerian description of the turbulent velocity and water vapor fields with a Lagrangian ensemble of cloud water droplets that can grow and shrink by condensation and evaporation, respectively.  The focus is on detailed analysis of the relaxation process of the droplet ensemble during the entrainment of subsaturated air, in particular the dependence on turbulence time scales, droplet number density, initial droplet radius and particle inertia. We find that the droplet evolution during the entrainment process is captured best by a phase relaxation time that is based on the droplet number density with respect to the entire simulation domain and the initial droplet radius. Even under conditions favoring homogeneous mixing, the probability density function of supersaturation at droplet locations exhibits initially strong negative skewness, consistent with droplets near the cloud boundary being suddenly mixed into clear air, but rapidly approaches a narrower,  symmetric shape.  The droplet size distribution, which is initialized as perfectly monodisperse, broadens and also becomes somewhat negatively skewed. 
Particle inertia and gravitational settling lead to a more rapid initial evaporation, but ultimately only to slight depletion of both tails of the droplet size distribution. 
The Reynolds number dependence of the mixing process remained weak over the parameter range studied, most probably due to the fact that the inhomogeneous mixing regime could not be fully accessed when phase relaxation times based on global number density are considered.
\end{abstract}
\maketitle

\section{Introduction}
\label{intro}
Mixing of a passive scalar in a turbulent flow is an archetypical problem in the study of turbulence and intermittency.  The mixing of cloudy and clear air adds additional complexity, not only because the scalar fields (e.g., droplet number density, water vapor concentration, and temperature) are no longer `passive' because of latent heating effects, but also because the condensed and vapor phases are coupled through mass conservation, and because the condensed phase itself can respond through various pathways.  For example, upon mixing of cloudy and clear air, droplets will evaporate until the mixture becomes saturated (assuming the initial presence of sufficient condensed water), but this could occur by all droplets evaporating by the same amount, or by a subset of droplets evaporating completely, leaving the remaining droplets unchanged.  The consequences for cloud properties are significant: droplet collision rates and cloud optical properties depend strongly on the shape of the droplet size distribution and the droplet number density.

The entrainment of clear air and its mixing with cloudy air occurs during the entire life
of a cloud. It introduces strong inhomogeneities at spatial scales ranging from 10$^2$ {\rm m}
down to 1{\rm mm} and at time scales from hours to seconds \cite{Blyth1993}. Through dilution, evaporation, and perhaps enhanced collision, the entrainment and mixing process changes the water droplet size distribution, which is of direct consequence to rain formation and cloud radiative properties and will eventually determine the cloud lifetime  itself~\cite{Shaw2003}. At stratocumulus cloud top, for example, entrainment and evaporation influence the entire cloud dynamics~\cite{Mellado2010}. Here, the presence of strong wind shear can additionally enhance the entrainment rate
\cite{Wang2008,Katzwinkel2011}. The mixing of clear and cloudy air can be characterized by the Damk\"ohler number, the ratio of a fluid time scale to a characteristic thermodynamic time scale associated with the evaporation process (phase relaxation time)
\begin{equation}
Da=\frac{\tau_{fluid}}{\tau_{phase}}\,.
\label{dam1}
\end{equation}
The two limits, $Da\ll 1$ and $Da\gg 1$ characterize the {\em homogeneous} and
{\em inhomogeneous} mixing, respectively. This notion was inspired by early laboratory experiments and the analogy with reacting flows \cite{Latham1977,Baker1984,Jensen1989}: Homogeneous mixing occurs when the condensational growth or evaporation of cloud water droplets is slow compared to the mixing time, and therefore takes place in a well-mixed environment; Inhomogeneous mixing occurs when the evaporation proceeds much faster than the flow structures evolve.  Both processes can coexist in a turbulent cloud since a whole spectrum of fluid time scales is present \cite{Lehmann2009}.

The aim of the present work is to gain a deeper understanding of the initial evolution of the mixing processes in a small subvolume at the clear air-cloud interface and to characterize the multiple-scale mixing processes. Specifically, we investigate how the droplet size distribution evolves as the mixing progresses. We conduct therefore a series of three-dimensional (3D) direct numerical simulations (DNS) which combine the Eulerian description of continuum fields such as velocity and vapor content with the Lagrangian evolution of an ensemble of cloud water droplets. In particular, we want to study the relaxation of the condensed phase during the entrainment process as a function of the initial droplet radius and the number density of the droplets. These two variables determine the phase relaxation time scale, and therefore influence the relative length and time scales at which homogeneous and inhomogeneous mixing predominate. Furthermore, we study the impact of turbulence on the entrainment and mixing by conducting simulations at different Reynolds number for the same initial vapor and droplet configuration. Finally, the effect of droplet inertia and gravitational settling is investigated. In order to focus on the one-way response of the droplet field to the turbulent mixing, we study the mixing processes in a simplified setting of model equations: the temperature is held fixed at a reference value of $T_0$ in the simulation domain, thus leaving the saturation vapor mixing ratio constant. As a consequence, the turbulence is not driven by buoyancy effects as in similar studies \cite{Vaillancourt2001,Andrejczuk2004,Andrejczuk2006}, but by a volume forcing that mimics a cascade of kinetic energy from larger scales obeying realistic amplitudes of the turbulent fluctuations that match the field measurements with ACTOS platform in \cite{Lehmann2009}. The reason for this choice is to disentangle the role of different processes on the cloud water droplet dynamics. The focus on Lagrangian behavior extends the prior computational studies of cloud mixing  \cite{Jensen1989,Andrejczuk2004,Andrejczuk2006}, allowing questions of variability in droplet growth history and droplet inertia to be directly addressed.  The Lagrangian perspective has been taken in other cloud studies \cite{Lanotte2009,Vaillancourt2001}, primarily with an emphasis on the bulk dynamics in a turbulent cloud and the resulting supersaturation field and droplet response (e.g., advection-diffusion equation for the supersaturation field). This study is focused rather on the microphysical response to a transient mixing event.  We continue in the following subsections by considering the time scales that are thought to govern the nature of that transient response.

\paragraph{Fluid time scale} Turbulent flows are characterized by a continuous range of time scales that can be associated with differently sized vortex structures or shear layers present in the flow. The largest time scale is the large scale eddy turnover time $T=L_{int}/u_{rms}$ where $L_{int}$ denotes a characteristic large (energy injection) scale of the flow and $u_{rms}$ is the root-mean square of the turbulent velocity fluctuations. The smallest mean time scale is the Kolmogorov time $\tau_{\eta}=\sqrt{\nu/\langle
\varepsilon\rangle}$ with the kinematic viscosity $\nu$ and the mean kinetic energy dissipation rate
$\langle\varepsilon\rangle$. For constant $\tau_{phase}$ the spectrum of possible Damk\"ohler numbers thus spans a range
\begin{equation}
Da_{\eta}=\frac{\tau_{\eta}}{\tau_{phase}}\ll Da \ll Da_L=\frac{T}{\tau_{phase}}\,.
\label{dam2}
\end{equation}
The crossover from homogeneous to inhomogeneous mixing is expected to be present at $Da\sim1$
and this can be associated with a length scale in the turbulent cloud \cite{Lehmann2009}.
Together with $\tau_{\ell}=\ell/v_{\ell}=\ell^{2/3}/\langle\varepsilon\rangle^{1/3}$ one gets
\begin{equation}
Da\sim 1\;\;\;\;\Leftrightarrow\;\;\;\; \ell_c\sim \sqrt{\langle\varepsilon\rangle\tau^3_{phase}}\,.
\label{dam2}
\end{equation}
In our simulations with constant $\langle\varepsilon\rangle$ it therefore follows that the phase relaxation determines the spatial transition from inhomogeneous mixing $(Da \gg 1)$ at large scales, to homogeneous mixing $(Da \ll 1)$ at small scales. Looked at from a somewhat different perspective, we find that the transition scale $\ell_c$ relative to the Kolmogorov length scale is simply related to a power of $Da_{\eta}$:
\begin{equation}
\frac{\ell_c}{\eta} \simeq \left( \frac{\tau_{phase}}{\tau_\eta} \right)^{3/2} \simeq  Da_\eta^{-3/2}\,.
\label{dam2}
\end{equation}

\paragraph{Phase relaxation time scale} The phase relaxation time is the exponential time scale associated with the condensational growth or evaporation of a population of droplets \cite{Lamb2011}. We provide a derivation here because it is not necessarily familiar within the turbulence community, yet it is of central importance in the cloud mixing problem. We have tried to simplify the derivation sufficiently that the key assumptions  are explicitly stated, but are not obscured by unnecessary details; in this regard the reader is also referred to Kostinski's lucid treatment \cite{Kostinski2009}. We begin by considering a single cloud droplet, assumed to be in equilibrium with its surrounding vapor field, so a steady mass flux of vapor toward the droplet surface is accompanied by a steady flux of latent heat away from the droplet surface \cite{Lamb2011,Rogers1989}.
Both steady fluxes are given by Fick's law
\begin{equation}
F_v=-D\frac{\mbox{d}\rho_v}{\mbox{d}R}\,,\;\;\;\;\;\;F_Q=-k \frac{\mbox{d}T}{\mbox{d}R}\,,
\label{flux1}
\end{equation}
with the vapor mass density $\rho_v$, the mass diffusivity $D$, the temperature $T$, and the
thermal conductivity $k$. At the droplet radius $r$, we define boundary conditions $T(R=r)=T_{r}$ and $\rho_v(R=r)=\rho_{v,r}$. In steady state the resulting profiles reach their asymptotic values of a reference temperature $T(R\gg r)=T_{\infty}$ and a reference vapor mass density $\rho_v(R\gg r)=\rho_{v,{\infty}}$  at several droplet radii.  Later it will be assumed that the reference values are the same for all droplets in a simulation `grid box,' which is an implicit statement that fluctuations in the vapor and temperature fields due to 'local' droplet interactions are neglected \cite{Kostinski2009}.  A steady regime requires energy conservation, i.e.,
\begin{equation}
F_v + L F_Q=0\,,
\label{flux2}
\end{equation}
with $L$ being the latent heat of vaporization (see Table 1).  Inserting Eq. (\ref{flux1}) into balance (\ref{flux2})
and integrating with respect to $R$ results in a relation between mass density values and temperatures
\begin{equation}
\frac{\rho_{v,{\infty}}-\rho_{v,r}}{T_r-T_{\infty}}=\frac{k}{LD}\,.
\label{flux3}
\end{equation}
It is further assumed that the water vapor pressure at the droplet surface is at the saturation
value $e_s(T_r)$ and thus $\rho_{v,r}=\rho_{vs}$ . Both are connected via the ideal
gas law $e_s=R_v \rho_{vs} T$ where $R_v$ is the vapor gas constant given in Table
1. By translating the linearized solution of the saturation pressure from the Clausius-Clapeyron
equation into an expression for the saturation vapor mass density, ones arrives together
with $T_r\approx T_{\infty}$ and (\ref{flux3}) at
\begin{equation}
\rho_{vs}(T_{\infty})-\rho_{vs}(T_r)\simeq\frac{L \rho_{vs}(T_{\infty})}{R_v T_{\infty}^2}(T_{\infty}-T_r)
=\frac{L \rho_{vs}(T_{\infty})}{R_v T_{\infty}^2} \frac{LD}{k}(\rho_{v,r}-\rho_{v,\infty})\,.
\label{flux4}
\end{equation}
The vapor flux across a sphere of radius $R$ has to be equal to the change of liquid water mass
inside the sphere
\begin{equation}
4\pi R^2 D \frac{\mbox{d}\rho_v}{\mbox{d}R}=\frac{\mbox{d}M_l}{\mbox{d}t}=4\pi\rho_l r^2\frac{\mbox{d}r}
{\mbox{d}t}\,.
\label{flux5}
\end{equation}
Integration from $R=r$ to $R=\infty$ results to
\begin{equation}
r \frac{\mbox{d}r}{\mbox{d}t}=\frac{D}{\rho_l}(\rho_{v,{\infty}}-\rho_{v,r})\,.
\label{flux6}
\end{equation}
With Eq. (\ref{flux4}), we can substitute $\rho_{v,r} (=\rho_{vs}(T_r))$ in Eq. (\ref{flux6}) and get the following equation for the radius growth by condensation
\begin{equation}
r \frac{\mbox{d}r}{\mbox{d}t}\simeq\frac{D}{\rho_l}\,(\rho_{v,{\infty}}-\rho_{vs}(T_{\infty}))
\left(1+\frac{D L^2 \rho_{vs}(T_{\infty})}{k R_v T_{\infty}^2}\right)^{-1} \simeq  \frac{{\cal D}}{\rho_l}\,
(\rho_{v,{\infty}}-\rho_{vs}(T_{\infty}))\,.
\label{flux7}
\end{equation}
The diffusivity constant ${\cal D}$ now incorporates the self-limiting effects of latent heat release This modified diffusivity can be written in terms of a ratio of two heat fluxes ${\cal D}/D = (1+\Phi_L/\Phi_k)^{-1}$: A characteristic heat flux due to latent heating resulting from a small change in droplet temperature, $\Phi_L = L D \Delta \rho_{vs}$, where $\Delta \rho_{vs}$ is obtained from $\Delta T$ through the linearized Clausius-Clapeyron equation; And a heat flux due to thermal conduction for the same temperature difference, $\Phi_k = k \Delta T$.  For typical warm cloud conditions $\Phi_L/\Phi_k$ is of order unity, so the heat-transfer-limited diffusivity ${\cal D}$ can be reduced by a factor of 2 (as is the case for the example values in Table 1). Only at rather low temperatures, e.g. $< -20$ $^\circ$C, do the thermal effects become negligible for liquid water.

The phase relaxation time scale is a direct consequence of the combination of
Eqns. (\ref{flux5}) and (\ref{flux7})
\begin{equation}
\frac{\mbox{d}M}{\mbox{d}t}=4\pi {\cal D} r (\rho_{v,{\infty}}-\rho_{vs}(T_{\infty}))\,,
\label{flux9}
\end{equation}
and the conservation of water mass, $n_d \mbox{d}M/\mbox{d}t=-\mbox{d}\rho_{v,{\infty}}/\mbox{d}t$. Here
$n_d$ is the droplet number density and thus
\begin{equation}
\frac{\mbox{d}\rho_{v,{\infty}}}{\mbox{d}t} = - 4\pi n_d {\cal D} r (\rho_{v,{\infty}}-\rho_{vs}(T_{\infty}))\,.
\label{flux10}
\end{equation}
If $T_{\infty}$ is constant, such that $\rho_{vs}(T_{\infty})$ is constant, this equation describes an exponential relaxation with a characteristic time constant, the phase relaxation time, given by
\begin{equation}
\tau_{phase}=\frac{1}{4\pi n_d {\cal D} r}\,.
\label{tauphase}
\end{equation}
It should be noted that a more detailed derivation expresses the phase relaxation time in terms of the first moment of the droplet size distribution (integral radius). Furthermore, we specifically call attention to the assumption of uniform number density, which of course is not exactly the case during an inhomogeneous mixing event in which the local number density varies considerably.  This is one of the motivations for taking a Lagrangian perspective, where individual droplets are followed through the flow, as opposed to treating the condensed phase as a continuous medium \cite{Andrejczuk2004,Andrejczuk2006}.

The outline of the manuscript is as follows. After introducing the set of equations which are solved
numerically in the present Euler-Lagrangian model, we will discuss in brief some properties of the
statistically stationary turbulent state. This discussion is followed by a description of the initial vapor content profile. The entrainment process is described afterwards in combination with an analysis
of the mean volume radius, the size distribution of the droplets and the supersaturation along the
droplet trajectories. Finally, we will add the effect of particle inertia and gravitational settling to the simulations
and quantify their impact. We conclude the work with a summary and an outlook.

\section{Model equations and numerical method}
\label{sec:1}
The turbulent velocity field ${\bf u}({\bf x},t)$ and the pressure field $p({\bf x},t)$ are those
necessary for the description of an incompressible turbulent flow. In this flow the vapor mixing
ratio field $q_v({\bf x},t)$ is transported and diffuses. The vapor mixing ratio is defined as
\begin{equation}
q_v({\bf x},t)=\frac{\rho_v}{\rho_d}\,,
\end{equation}
where $\rho_v$ and $\rho_d$ are the mass densities of vapor and dry air, respectively. For the purposes of this paper, the advection-diffusion equation for temperature is not considered.

The Eulerian equations for the turbulent fields are
\begin{eqnarray}
{\bf \nabla}\cdot{\bf u}&=&0\,,
\label{euler1}\\
\partial_t {\bf u} + ({\bf u}\cdot{\bf\nabla}) {\bf u}  &=&-\frac{1}{\rho_0}{\bf\nabla}p+
\nu{\bf\nabla}^2{\bf u}+{\bf f}\,,
\label{euler2}\\
\partial_t q_v + {\bf u}\cdot{\bf\nabla} q_v & = & D
{\bf\nabla}^2 q_v -C_d\,,
\label{euler4}
\end{eqnarray}
where ${\bf f}({\bf x},t)$ is a bulk forcing which sustains the turbulence and
$C_d$ is the condensation
rate. The entrainment is studied in a cube with volume $V=L_x^3$ and with periodic
boundary conditions in all three spatial directions.  It is spanned by an equidistant mesh
with $N_x^3$ cells of mesh size $a$. The Eulerian equations are solved by a pseudospectral
method using fast Fourier transformations. Time advancement is done by a second-order
predictor-corrector method. The spectral resolution in the present cases is $k_{max}\eta=3$
with the maximum resolved wavenumber $k_{max}=2\pi\sqrt{2}N_x/(3 L_x)$ and the Kolmogorov
scale $\eta$ (see Table 1). Grid sizes used  throughout this work are $N_x^3=128^3,\,256^3$ and $512^3$ corresponding with turbulent flows at Taylor microscale Reynolds numbers
$R_{\lambda}=42,\,59$ and 89, respectively.

\paragraph{Volume forcing} Here, we consider a turbulent flow that is sustained by a volume
forcing ${\bf f}({\bf x},t)$ in a statistically stationary turbulent state. This driving is implemented
in the Fourier space for some modes with the smallest wavenumbers $k_f$ only, i.e. $k^{-1}_f
\approx L_x$. The kinetic energy is injected at a fixed rate $\epsilon_{in}$ into the flow. The
volume forcing is established by the expression \cite{Schumacher2007}
\begin{eqnarray}
{\bf f}({\bf k},t)&  =  & \epsilon_{in} \frac{{\bf u}({\bf
k},t)}{\sum_{{\bf k}_f\in {\cal K}} |{\bf u}({\bf k}_f,t)|^2}\,\delta_{{\bf k},{\bf k}_f}\,,
\label{fordef}
\end{eqnarray}
with the Kronecker delta
\begin{equation}
\delta_{{\bf k},{\bf k}_f} = 1\;\;\;\; \mbox{if} \;\;\;{\bf k} = {\bf k}_f, \;\;\;\;\;
\delta_{{\bf k},{\bf k}_f} = 0\;\;\;\; \mbox{otherwise}\,,
\end{equation}
and the wavevector subset ${\cal K}$ which contains some wave vectors, e.g. ${\bf k}_f=(1,1,2)$
plus all permutations with respect to components and signs. Since the large-scale velocity follows a Gaussian statistics, the present forcing will act in similar way as other stochastic forcing schemes \cite{Eswaran1988}.  The present energy injection mechanism prescribes the mean energy dissipation rate; that is, the magnitude of the first moment of the energy dissipation rate field, $\langle\epsilon\rangle$, is determined by the injection rate, $\epsilon_{in}$, having no Reynolds number dependence. This can be seen as follows. Given the periodic boundary conditions in our system, the turbulent kinetic energy balance, which results from rewriting (\ref{euler2}) in the Fourier space, follows to
\begin{equation}
\frac{\mbox{d} E_{kin}}{\mbox{d}t}=-\nu \sum_{\bf k} k^2 |{\bf u}({\bf k},t)|^2 +
\sum_{\bf k} {\bf f}({\bf k},t) {\cdot \bf u}^{\ast}({\bf k},t)\, ,
\label{balance1}
\end{equation}
where ${ \bf u}^{\ast}$ is the conjugate complex Fourier mode. The first term on the right hand side of (\ref{balance1}) is the volume average of the energy dissipation rate. Additional time averaging in combination with (\ref{fordef}) results in
\begin{equation}
\nu \sum_{\bf k} k^2 \langle|{\bf u}({\bf k},t)|^2\rangle_t=\langle\epsilon\rangle=\epsilon_{in}=
\sum_{\bf k} \langle{\bf f}({\bf k},t) {\cdot \bf u}^{\ast}({\bf k},t)\rangle_t\,.
\label{epsiloninput}
\end{equation}
The applied driving thus allows a full control of the mean energy dissipation rate
$\langle\epsilon\rangle$ and thus the Kolmogorov scale $\eta$ via the parameter $\epsilon_{in}$
in (\ref{fordef}).
\begin{table}[t]
\caption{List of constants and reference values are given in the upper part of the table. Simulation parameters and characteristics of the statistically stationary turbulent state follow in the second one.}
\centering
\label{tab:1}
\begin{tabular}{llll}
\hline
Quantity & Symbol & Unit & Value  \\
\hline
Reference temperature    & $T_{\infty}$ & $K$     & 270\\
Reference pressure          & $p_{\infty}$ & $hPa$ & 845 \\
Kinematic viscosity            & $\nu$ & $m^2 s^{-1}$ & $1.5\times 10^{-5}$\\
Vapor diffusivity at $T_{\infty}$ &$D$ & $m^2 s^{-1}$ &$2.16\times 10^{-5}$\\
Modified vapor diffusivity at $T_{\infty}$ &${\cal D}$ &$m^2 s^{-1}$ &$1.31\times 10^{-5}$\\
Thermal  conductivity of air at $T_{\infty}$ & $k$ & $J\,m^{-1}s^{-1}K^{-1}$ & $2.38\times 10^{-2}$\\
Gravity acceleration & $g$  & $m\,s^{-2}$ & $9.81$\\
Gas constant for water vapor & $R_v$ & $J\,K^{-1}\,kg^{-1}$ & $461.5$\\
Gas constant for dry air & $R_d$ & $J\,K^{-1}\,kg^{-1}$ & $287.0$\\
Specific heat at constant pressure & $c_p$ & $J\,kg^{-1} K^{-1}$ & $1005$\\
Latent heat & $L$ & $J\,kg^{-1}$ & $2.5\times 10^6$\\
Liquid water density & $\rho_l$ & $kg\,m^{-3}$ & $10^3$\\
Reference mass density of air & $\rho_0$ & $kg\,m^{-3}$& $1.06$\\
Saturation pressure at $T_{\infty}$ & $e_s(T_{\infty})$ & $Pa$ & $484$\\
Saturation vapor density at $T_{\infty}$ & $\rho_{vs}(T_{\infty})$ & $kg\,m^{-3}$ & $3.9\times 10^{-3}$\\
Constant in Eq. (\ref{flux8})& $K$ & $m^2\,s^{-1}$ & $5.07\times 10^{-11}$\\ \hline
Box length  & $L_x$ & $m$ & $0.128,\;0.256,\;0.512$\\
Grid resolution & $a$ & $mm$ & $1.0$\\
Kolmogorov scale & $\eta=\nu^{3/4}/\langle\varepsilon\rangle^{1/4}$ & $mm$ & $1.0$\\
Mean energy dissipation rate & $\langle\varepsilon\rangle\;(=\varepsilon_{in})$ & $m^2 s^{-3}$& $0.003375$\\
Root-mean-square velocity & $u_{rms}$ & $cm\,s^{-1}$ & $8.6,\; 10.1,\; 12.5$\\
Taylor microscale Reynolds number &
$R_{\lambda}=\sqrt{5/(3\nu\langle\varepsilon\rangle)}\,u_{rms}^2$ & & $42,\; 59,\; 89$\\
Initial cloud water droplet radius & $R_0=r(t=0)$ & $\mu m$& $10,\; 15,\; 20$\\
Cloud number density & $n_d$ & $cm^{-3}$ &  $62,\; 82,\; 164,\; 328$\\
\hline
\end{tabular}
\end{table}

\paragraph{Cloud water droplet advection and condensation rate}
The Lagrangian evolution of  each of the $N$ droplets in the volume $V$ is described by
the following set of equations
\begin{eqnarray}\label{eqn:Lag_position}
\frac{\mbox{d}{\bf X}(t)}{\mbox{d}t}&=&{\bf V}(t)\,,
\label{lag1}\\
\frac{\mbox{d}{\bf V}(t)}{\mbox{d}t}&=&\frac{1}{\tau_p}[{\bf u}({\bf X},t)-{\bf
V}(t)]+{\bf g}\,.
\label{lag2}
\end{eqnarray}
Here, ${\bf X}$ is the droplet position and ${\bf V}$ its velocity. We consider both droplets
which match perfectly with the surrounding fluid velocity as well as inertial particles with a finite
particle response time $\tau_p=2\rho_l r^2/(9 \rho_0 \nu)$.

As droplets are advected by the fluid they can grow or evaporate in response to the local vapor field (recalling that in this study temperature is constant). Direct droplet interactions through collision are neglected in order to focus solely on the initial stage of the entrainment and mixing process. The vapor mixing ratio can be coupled to droplet growth by defining the supersaturation $S=\rho_{v,\infty}/\rho_{vs}(T_{\infty})-1$, such that
\begin{equation}
S({\bf x},t)=\frac{q_v({\bf x},t)}{q_{v,s}}-1\,.
\label{supersat}
\end{equation}
Then it follows from Eqn. (\ref{flux7}) in Sec. \ref{intro} that the droplet growth rate can be written as
\begin{equation}
r \frac{\mbox{d}r}{\mbox{d}t}=KS \;\;\;\;\;\;\;\;\; \mbox{with}\;\;\;\;\;\;\;\;\;
K= \left[ \rho_l \left(\frac{R_v T_{\infty}}{D e_s(T_{\infty})}+\frac{L^2 }{k R_v T_{\infty}^2}\right)\right]^{-1}=\frac{\rho_{vs}(T_{\infty})}{\rho_l}{\cal D}\,.
\label{flux8}
\end{equation}
In the Lagrangian frame this condensational growth process becomes
\begin{equation}\label{eqn:Lag_radius}
r(t)\frac{\mbox{d}r(t)}{\mbox{d}t}= K S({\bf X},t)\,.
\end{equation}
We calculate the condensation rate field $C_d({\bf x},t)$ following \cite{Vaillancourt2001} by
\begin{equation}
C_d({\bf x},t)=\frac{1}{m_a} \frac{\mbox{d}m_l({\bf x},t)}{\mbox{d}t}
=\frac{4\pi\rho_l K}{\rho_0 a^3} \sum^{\triangle}_{\beta=1}S({\bf X}_{\beta},t) r(t)\,,
\label{cond1}
\end{equation}
where $m_a$ is the mass of air per grid cell and the sum collects the droplets inside
each of the grid cells of size $a^3$ that surround the (grid) point ${\bf x}$.  This relation
closes the system of Eulerian-Lagrangian equations.  The transmission of the Eulerian field
values at grid positions  to the enclosed droplet position is done by trilinear interpolation. The
inverse procedure is required for the calculation of the condensation rate which is evaluated
at first at the droplet position and then redistributed to the nearest eight grid vertices.
\begin{figure}
\centering
\includegraphics[width=0.9\textwidth]{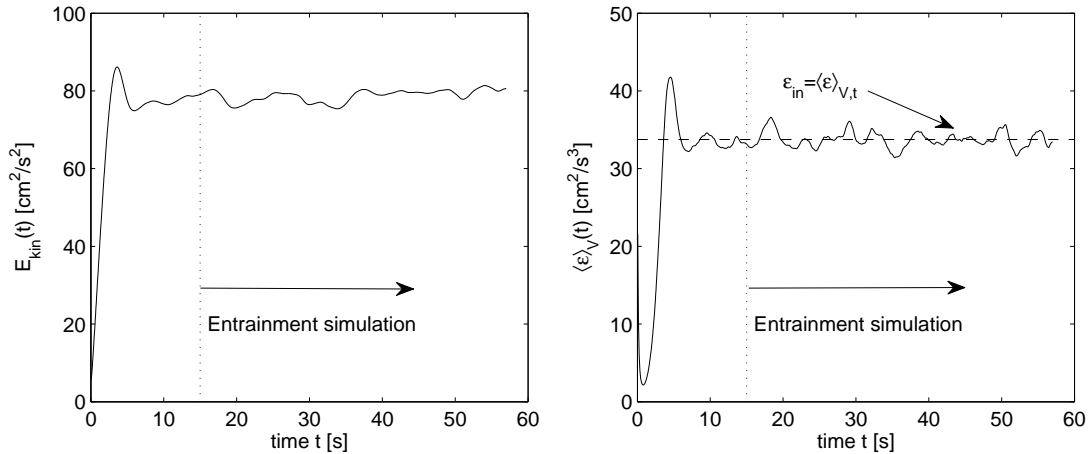}
\caption{Turbulent kinetic energy (left) and volume averaged kinetic energy dissipation rate (right) as a function of time. The dashed line in the right panel marks the prescribed ensemble average $\langle\varepsilon\rangle_{V,t}$ (see Table 1). For $t>15s$, both quantities and the turbulence as a whole are fully relaxed into a statistically stationary state. This is the starting point of the entrainment simulation which is marked by a vertical dotted line. Data are for the run at $R_{\lambda}=89$.}
\label{fig:1}
\end{figure}
\begin{figure}
\centering
\includegraphics[width=0.8\textwidth]{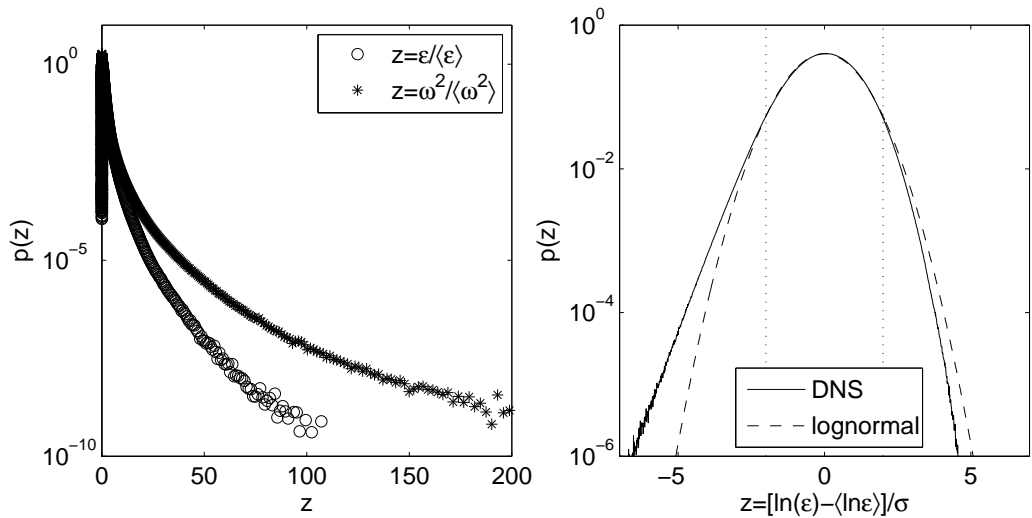}
\caption{Statistics of the velocity gradients. Left: Probability density functions (PDF) of the energy dissipation rate field $\varepsilon({\bf x},t)$ and the enstrophy density $\omega^2({\bf x},t)$. Right:
Comparison of the energy dissipation rate field statistics with the lognormal prediction of the refined similarity hypothesis \cite{Kolmogorov1962}. The dotted vertical lines are at $z=\pm 2$.
Data are again for the run at $R_{\lambda}=89$.}
\label{fig:2}
\end{figure}

The particle advection poses a numerical problem when particle inertia is considered. Evaporating droplets
cause a particle response time $\tau_p\to 0$ thus making Eqn. (\ref{lag2}) stiff. Therefore a
semi-implicit second-order particle advection scheme is chosen.  While the equations for the radius and
the droplet position are solved by a predictor-corrector scheme, the droplet velocity equations are
solved by a combination of an implicit forward Euler step that is required for the corrector step of the
droplet positions and a trapezoidal scheme for the velocity itself. We verified the accuracy of this
scheme by the analytical test case of the freely falling droplet with friction for ${\bf u}=0$ for Stokes
numbers down to $St_{\eta}=\tau_p/\tau_{\eta}\sim 10^{-4}$.

The complete list of thermodynamic reference values and constants is summarized in Table 1. The values for reference density, temperature, pressure and the resulting saturation values are chosen in agreement with recent airborne measurements by Lehmann et al. \cite{Lehmann2009}.  It is worth emphasizing again that, as discussed in the introduction, we have set up the model equations such that there is no active feedback to the turbulent dynamics through the temperature field. Specifically, while the effects of latent heat and thermal conductivity are included in the droplet growth rate (e.g., through the modified diffusivity ${\cal D}$ in Eqn. (\ref{flux7})), there is no coupling of the condensation rate (which is given by Eqn. (\ref{cond1})) to the momentum equation (\ref{euler2}) via a buoyancy term, and no advection-diffusion of a temperature field.  This study is focused on the vapor advection-diffusion aspects of the mixing problem.

\section{Preparation of the initial turbulence state}
In order to prepare the turbulence initial conditions for the Euler-Lagrangian simulations we first run a pure flow simulation with the volume driving described by (\ref{fordef}). Figure \ref{fig:1} demonstrates the relaxation into a statistically stationary state by means of the time traces of the turbulent kinetic energy (left) and the volume-averaged energy dissipation rate (right). For times larger than 15 seconds both quantities are found to fluctuate moderately about their temporal means. We also checked that the isotropy of the flow is established by comparing the mean squares of the three velocity components. Figure \ref{fig:2} shows the probability density function (PDF) of the energy dissipation rate field and the square of vorticity magnitude, denoted as enstrophy density, which are given by
\begin{equation}
\varepsilon({\bf x},t)=\frac{\nu}{2} \left(\frac{\partial u_i}{\partial x_j}+\frac{\partial u_j}{\partial x_i}\right)^2
\,,\;\;\;\;\;\;\; \omega^2({\bf x},t)=\left(\epsilon_{ijk}\frac{\partial u_k}{\partial x_j} \right)^2\,.
\label{energydiss}
\end{equation}
The stretched exponential tails of both quantities demonstrate the enhanced spatial intermittency of the
velocity gradients at the smaller scales (see e.g. \cite{Schumacher2010}).  The right panel displays the
PDF of the energy dissipation rate field (same data as in the left panel) in comparison with the refined
similarity hypothesis prediction by Kolmogorov \cite{Kolmogorov1962}. Deviations in both tails for
$|z|>2$ are found. Two aspects contribute, in our view, to the deviations. First, the Reynolds
number of the present simulations are still moderate. Second, our spectral resolution exceeds
standard resolutions by at least a factor of 2. In  Ref. \cite{Schumacher2007} it was demonstrated that
the higher spectral resolution is necessary to resolve the tails, i.e. the rare high-amplitude events,
sufficiently well.  We also verified from the statistical analysis that the relation $\langle\varepsilon
\rangle=\nu\langle\omega^2\rangle$ is satisfied.
\begin{figure}
\centering
\includegraphics[width=0.32\textwidth]{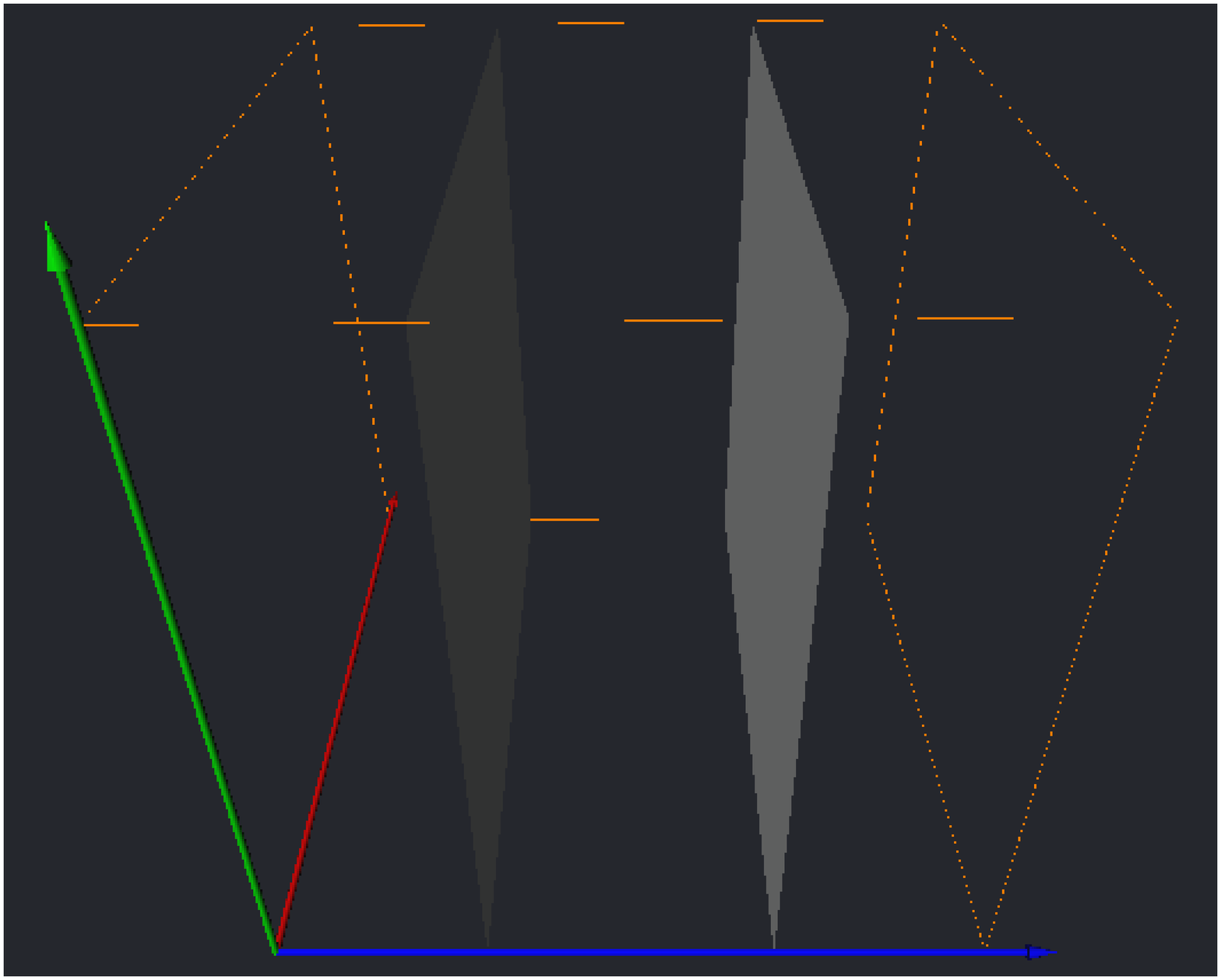}
\includegraphics[width=0.32\textwidth]{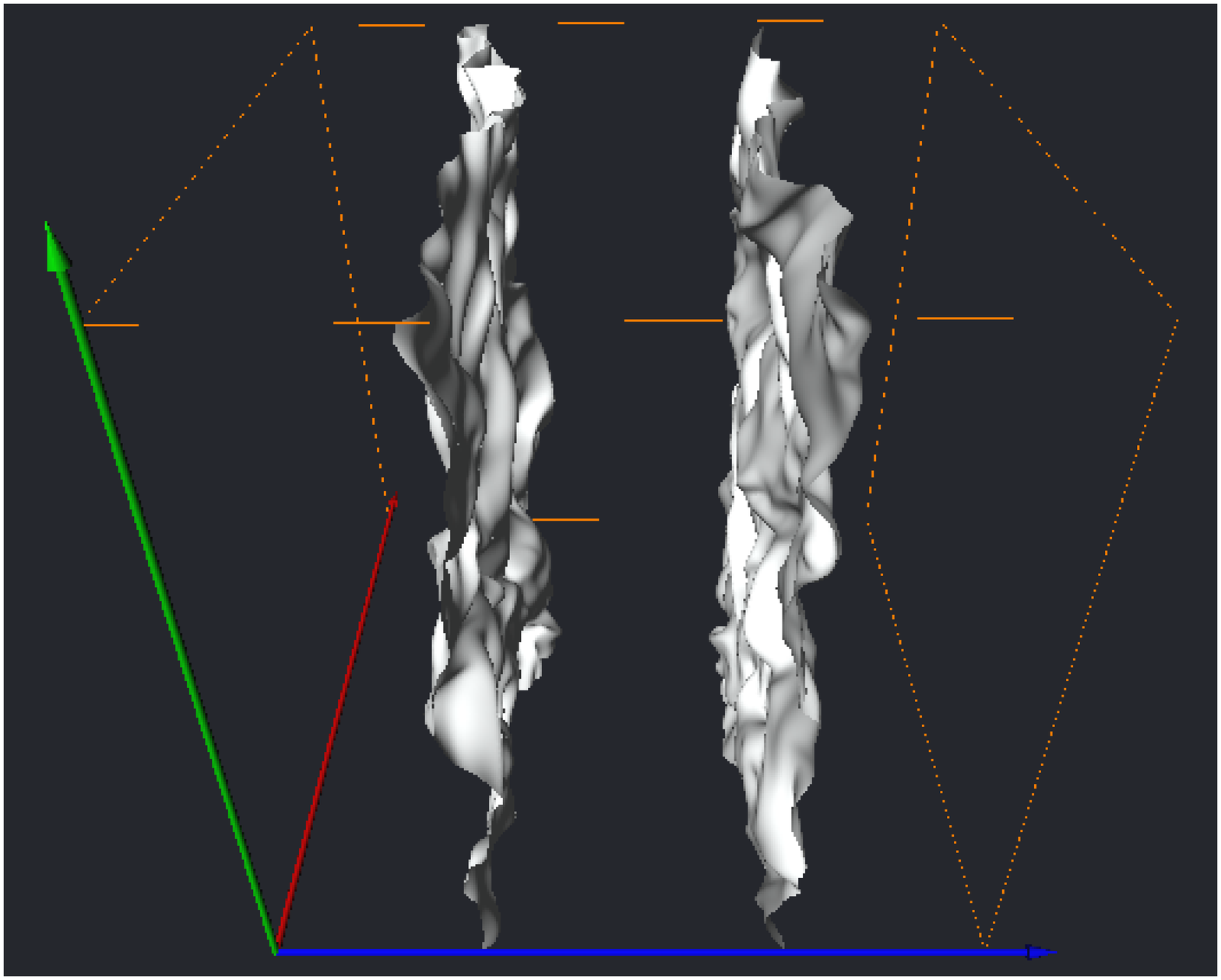}
\includegraphics[width=0.32\textwidth]{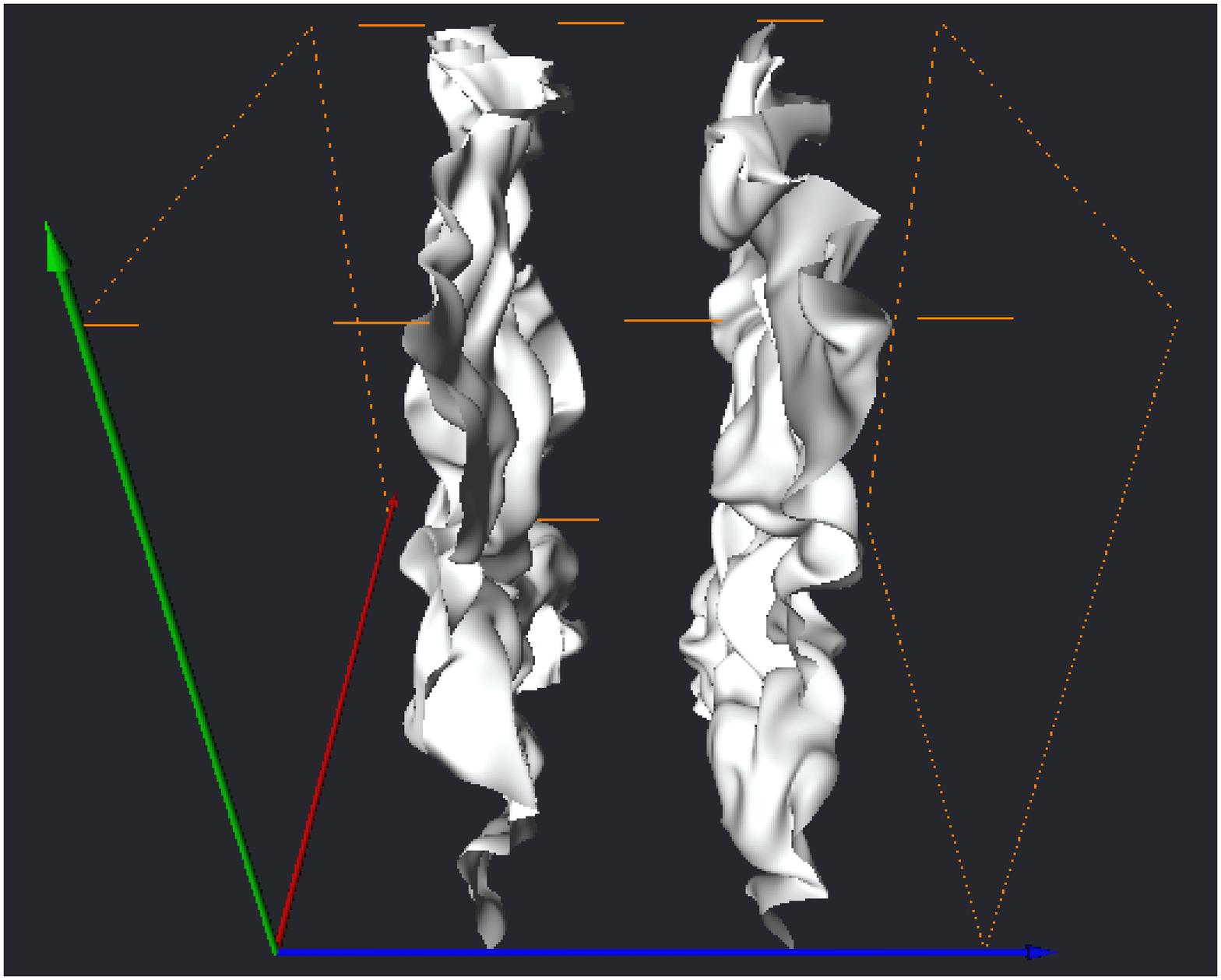}

\vspace{0.04cm}
\includegraphics[width=0.32\textwidth]{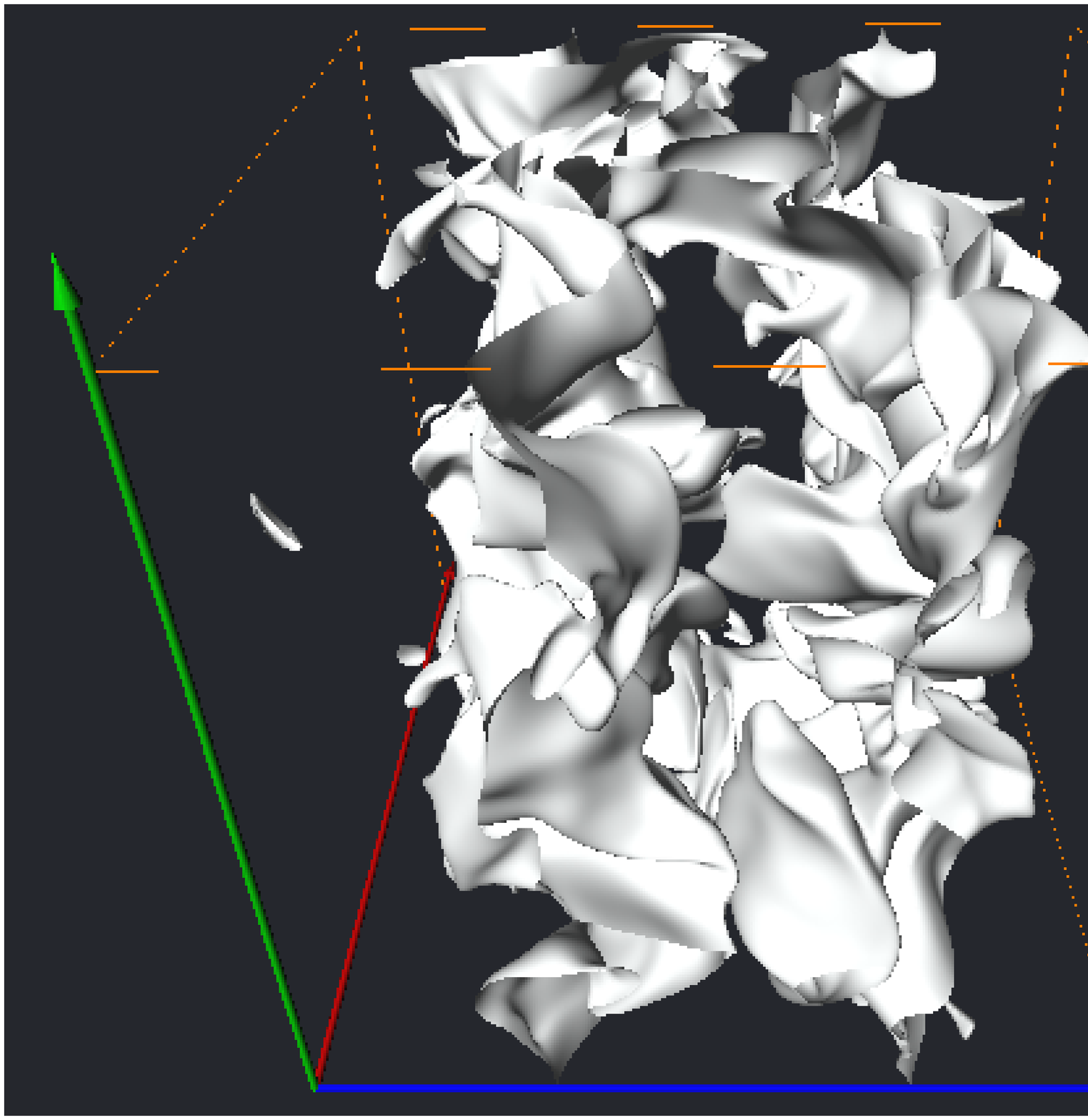}
\includegraphics[width=0.32\textwidth]{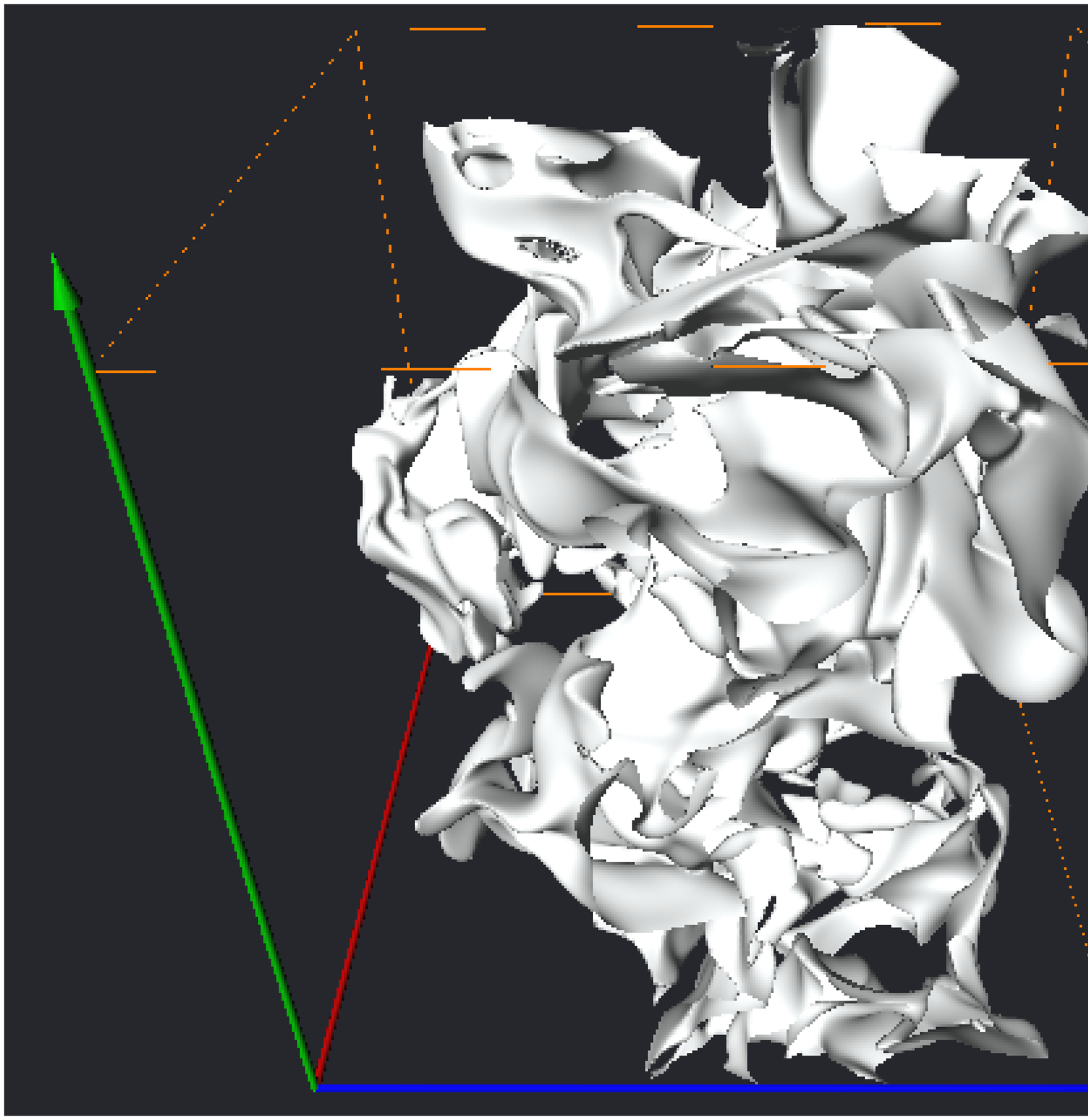}
\includegraphics[width=0.32\textwidth]{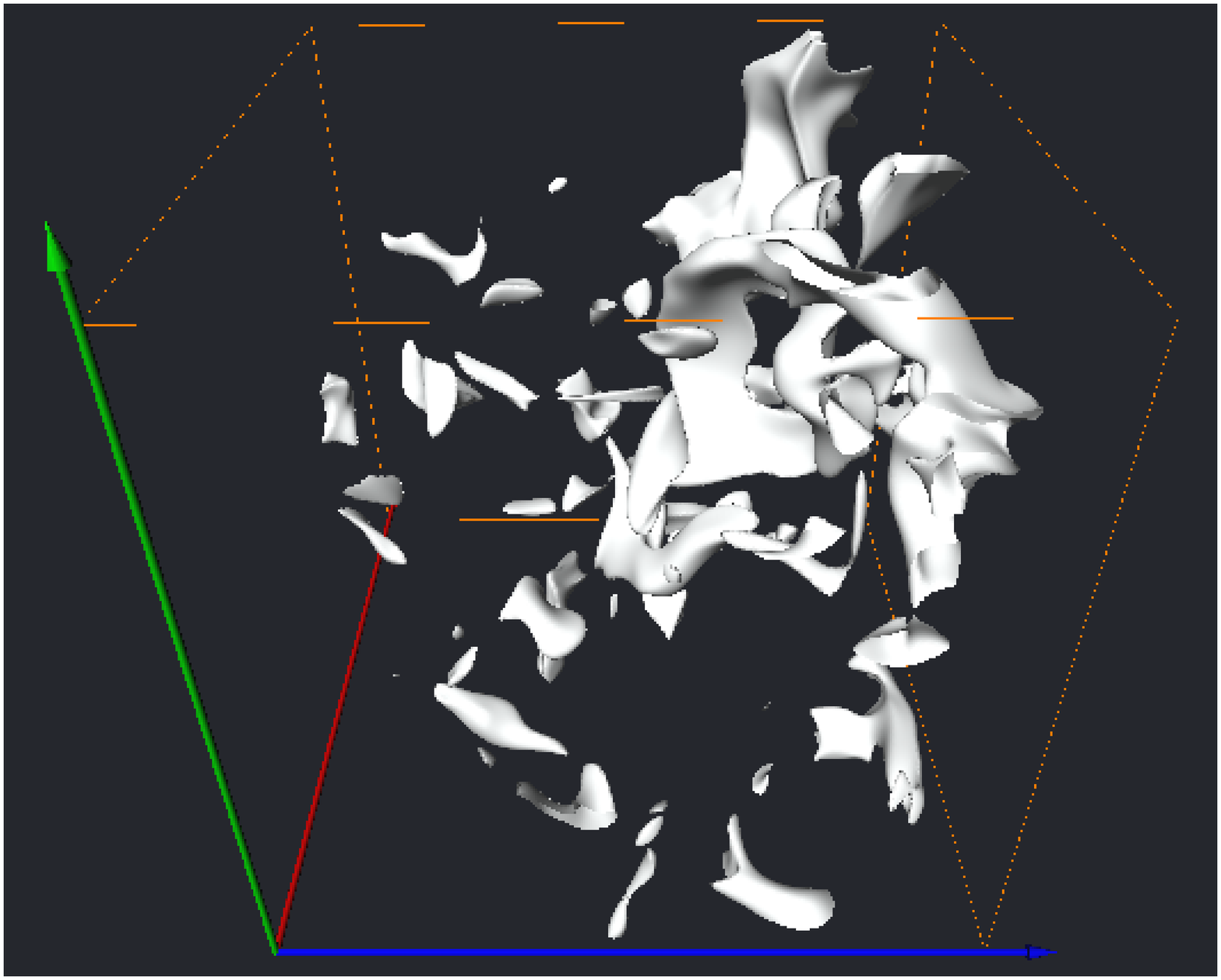}
\caption{(Color online) Time evolution of the isosurfaces at $q_v=q_{vs}$. From top left to bottom right: $t=0.0\,s$,
$t=0.1\,s$, $t=0.2\,s$, $t=0.6\,s$, $t=1.4\,s$ and $1.9\,s$. Droplets are seeded initially between the two vertical
planes of the top left figure. Data are for the run with $N_x^3=256^3$ at $R_{\lambda}=59$.}
\label{fig:3}
\end{figure}

In Table 1 we summarize the simulation parameters and turbulence quantities. The numbers are matched to typical magnitudes for turbulent cumulus clouds \cite{Lehmann2009}. The developed turbulent state serves as the initial condition for our entrainment investigations.

By definition, mixing presumes that the initial state possesses imposed gradients in the fields, but the form of those gradients is by no means obvious.  For example, the initial vapor-droplet configuration depends on assumptions regarding the degree of correlation between the two fields, as well as the spatial scales of the fluctuations.  In order to focus on mixing from a state possessing a single, clearly defined initial length scale, we have chosen to begin with an idealized slab-cloud geometry.  This slab could be considered analogous to a cloud edge or cloud filament boundary formed through inhomogeneous mixing at larger scales.  More precisely, the initial condition is a slab-like filament of supersaturated vapor that fills about one third of the cubic simulation box, with the vapor profile across the slab given by
\begin{equation}
q_v(x,y,z,t=0)=(q^{max}_{v}-q^{e}_{v}) \exp\left[-A(x-x_0)^6\right] + q^{e}_{v}\,.
\label{profile1}
\end{equation}
Here, $q^{max}_{v}$ is the maximum amplitude of $q_v$, which exceeds the saturation value $q_{vs}
(T_0)$ by 2\%. The variable $q^{e}_{v}$ stands for the environmental vapor mixing ratio, representing the subsaturated clear air outside the supersaturated filament. Droplets are seeded randomly in the supersaturated slab as a monodisperse initial ensemble. Technically, a sharp boundary between clear and cloudy air on a length scale smaller than $l_c$ is realistic only in a transient sense, and other vapor profiles have been considered (see e.g. \cite{Andrejczuk2004}).  Furthermore, as already stated, we use fully developed, forced turbulence throughout the volume, which is idealized with respect to the turbulence level in- and outside a cloud (see e.g. \cite{Knaepen2004}). One might consider, however, that both the transient sharp boundaries and the uniformity of turbulence energy dissipation rate mimics the lower end of the turbulent energy cascade during a mixing event.  Most importantly, the justification for the idealized initial conditions is that they allow us to study the entrainment in a well-controlled setup with clearly defined length and time scales, and therefore to disentangle the contributions of different physical processes to the droplet dynamics.

We proceed now to perform a series of simulations that combine three initial radii $R_0=r(t=0)$ with four different droplet number densities $n_d$
as listed in Tab. \ref{tab:2}.  Corresponding to these number densities ($n_d$), the total numbers ($N$) of advected droplets are $412.500,\,550.000,\,
1.100.000$ and $2.200.000$, respectively for $R_{\lambda}=59$. Hence, a total of 12 simulations are run for a time interval of $60$ seconds which corresponds with $60.000$ simulation time steps. In addition, we add two runs at different Reynolds numbers. All values have been chosen so as to be representative of a warm cumulus cloud, but also so as to allow exploration over a realistic range of microphysical time scales (e.g., phase relaxation time as defined in Eqn. (\ref{tauphase})).  The `global' number density $n_d^{(g,0)}$ and liquid water content $w^{(g,0)}$, i.e., the values obtained when droplet number and liquid water content are averaged over the entire computational domain, for all simulations are also listed in Tab. \ref{tab:2}.  Phase relaxation times based on both the initial cloud properties and the global averages are also listed, as well as the resulting Damk\"ohler numbers. These will be discussed in detail later, but for now suffice to say that cloud values of $Da_\eta$ are always less than 0.1, safely in the homogeneous mixing limit, and cloud values of $Da_L$ vary from approximately 0.3 to 3, thereby just making the transition toward inhomogeneous mixing at the large scales.  Several of the liquid water contents are rather high, but this allows the largest $Da_L$ to be achieved.
\begin{figure}
\centering
\includegraphics[width=0.8\textwidth]{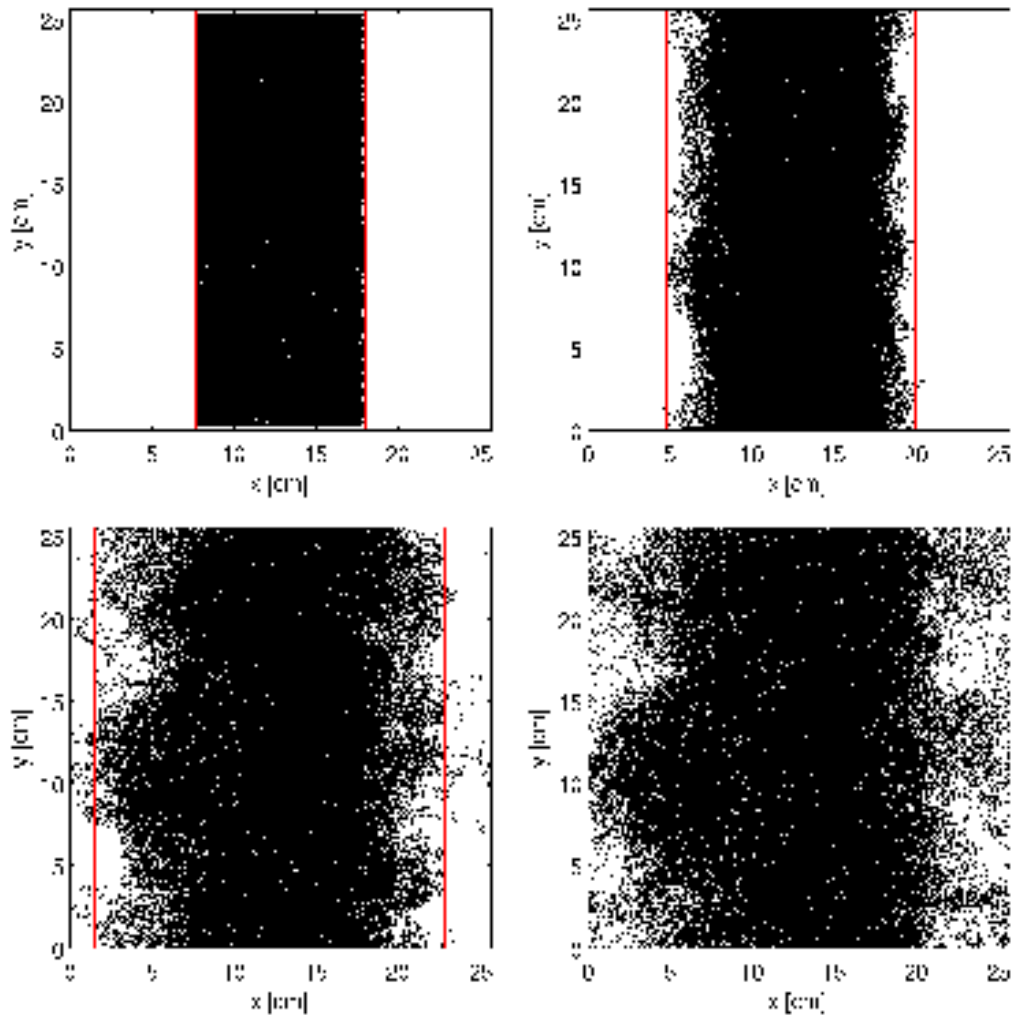}
\caption{(Color online) Time evolution of the droplet ensemble. From top left to bottom right: $t=0.0\,s$,
$t=0.2\,s$, $t=0.7\,s$ and $t=1.0\,s$. The positions of all $5.5\times 10^5$ (which corresponds with
$n_d=82 cm^{-3}$) are projected onto the $x$--$y$ plane. The vertical lines in the first three panels indicate the minimum/maximum $x$ position up to which vapor filaments with $q=q_{vs}$ have been
advected. Data are the same as in Fig. \ref{fig:3}.}
\label{fig:4}
\end{figure}
\begin{figure}
\centering
\includegraphics[width=0.6\textwidth]{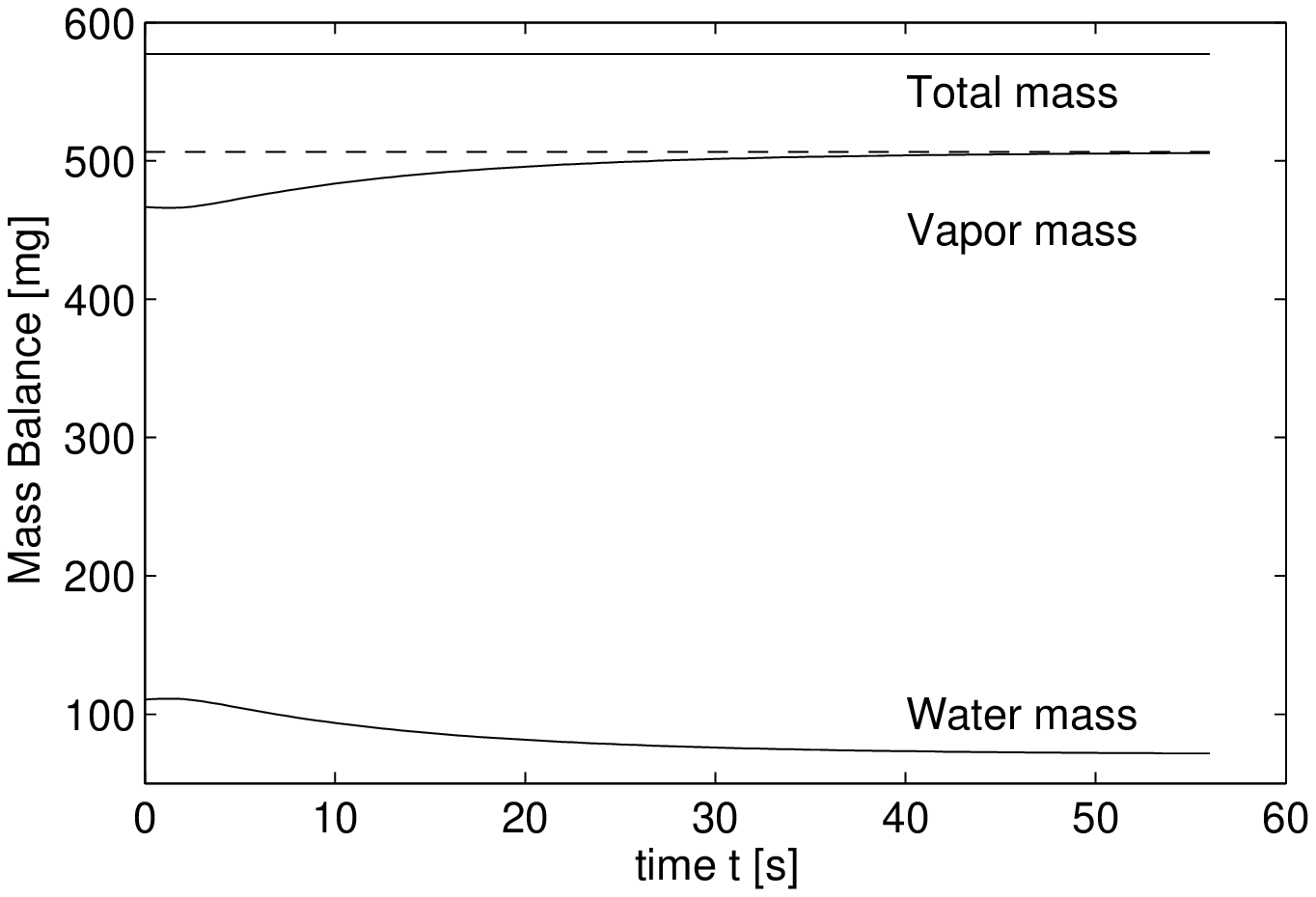}
\caption{Total water content conservation versus time. The total mass in volume $V$ consists
of the vapor mass $M_v$ and the liquid water mass $M_l$. The data are for the run at
$R_{\lambda}=89$. The saturation level and the corresponding $M_{vs}$ are indicated by the
dashed line.}
\label{fig:5}
\end{figure}

\section{Entrainment process and droplet evolution}
Figure \ref{fig:3} displays the time evolution of the turbulent entrainment process. The two isosurfaces are shown for which the vapor content satisfies $q_v(x,y,z,t)=q_{vs}$. Initially planar, they become highly convoluted as the time progresses. For times larger than 2 seconds, approximately one large eddy time, the vapor fluctuations decay below the saturation mixing ratio, leading to evaporation of all droplets. In Fig. \ref{fig:4} the same evolution is shown from the Lagrangian perspective. We project the position of the individual droplets of the whole ensemble onto the $x$--$y$ plane. The additional vertical lines indicate the minimum and maximum $x$ position up to which filaments with $q_v(x,y,z,t)=q_{vs}$ have been advected by the turbulent flow.
At the beginning of the evolution, droplets and vapor filaments follow the same stretching and twisting phenomena. With progressing time, however, both dynamics become decoupled as the vapor field is subjected to additional diffusion and a weak source-sink contribution from the condensation rate. This can be observed by comparison of the upper right ($t=0.2\,s$) and lower left ($t=0.7\,s$) panels.

Figure \ref{fig:5} shows the total water balance for a typical evolution run. The sum of vapor and liquid water masses in the volume $V$, $M_v$ and $M_l$, has to be constant with respect to time
\begin{equation}
M_v(t)+M_l(t)=\rho_0 \int_V q_v(x,y,z,t)\,\mbox{d}V + \frac{4}{3}\pi\rho_l\sum_{i=1}^{N_p}r^3_i(t)=const.
\label{balance}
\end{equation}
We can distinguish two cases for the dynamics of the present entrainment problem:
\begin{itemize}
\item Case 1 is present if
\begin{equation}
M_l(t=0)\le M_{vs}-M_v(t=0)\,,
\end{equation}
where $M_{vs}=\rho_0 q_{vs} V$ is the total mass of the saturated vapor. As a consequence
\begin{equation}
q_v\to \overline{q}_{v,\infty} \le q_{vs} \;\;\;\;\;\;\;\;\; \mbox{and}\;\;\;\;\;\;\;\;\;r_i\to 0\;\;\;\forall i=1...N\,.
\end{equation}
Eventually all droplets evaporate and the vapor content fluctuations decay. The vapor content
relaxes to a homogeneous field amplitude $\overline{q}_{v,\infty}$ below the saturation level.
\item Case 2 is present if
\begin{equation}
M_l(t=0)>M_{vs}-M_v(t=0)\,.
\end{equation}
In this scenario the dynamics ends with the state
\begin{equation}
q_v\to q_{vs} \;\;\;\;\;\;\;\;\; \mbox{and}\;\;\;\;\;\;\;\;\;r_i\to r_{i,\infty}\;\;\;\forall i=1...N\,.
\end{equation}
Eventually, the vapor content fluctuations decay to zero and the vapor content relaxes to the saturation value $q_{vs}$. Thus $S({\bf x},t)=0$ and
all droplet growth or evaporation ceases, leaving a droplet population with a mean volume diameter $\langle r^3\rangle_{\infty}$. It should further
be noted that $N$ is not necessarily conserved during the process because some subset of the droplet population may evaporate completely.
\end{itemize}
In the following Lagrangian statistical analyses of the droplet growth, the supersaturation along the droplet paths and the relaxation time scales are discussed. Due to the initial excess of saturation level, the majority of the droplets can grow in the first one to two seconds.  Subsequently they start to shrink slowly. The progressing entrainment of clear air generates local filaments of subsaturated vapor for an increasing number of droplets beginning with those closer to the original planar cloud-clear air interface.
\begin{table}[t]
\caption{List of parameters for the phase relaxation studies. We list the initial droplet radius $R_0$,
the {\em initial} number densities with respect to the slab cloud, $n_d^{(c,0)}$, and the whole volume,
$n_d^{(g,0)}$, the global liquid water content $w^{(g,0)}$ in $g\,m^{-3}$, the single-droplet evaporation
time $\tau_r$, the phase relaxation times (\ref{tauphase}) based on the number densities $n_d^{(c,0)}$
and $n_d^{(g,0)}$ , the numerically determined relaxation time $\tau_{relax}$, and the large-scale eddy
turnover time $T$. Furthermore the four possible Damk\"ohler numbers and the Taylor microscale Reynolds
number are given. All number densities are in $cm^{-3}$ and all times in $s$. The Kolmogorov time scale
in all 14 DNS runs is $\tau_{\eta}=0.067 s$. Empty entries for $\tau_{relax}$ stand for the complete
evaporation of the droplets (Case 1).}
\centering
\label{tab:2}
\begin{tabular}{llllllllllllll}
\hline\
$R_0$ & $n_d^{(c,0)}$ & $n_d^{(g,0)}$ & $w^{(g,0)}$ & $\tau_r$ & $\tau^{(c,0)}_{phase}$ & $\tau^{(g,0)}_{phase}$ & $\tau_{relax}$ & $T$ & $Da^{(c,0)}_{\eta}$ & $Da^{(c,0)}_L$ &
$Da^{(g,0)}_{\eta}$ & $Da^{(g,0)}_L$ & $R_{\lambda}$ \\
\hline
10 & 62   & 25  & 0.1 & 9.1 & 9.8  & 24.7  &  --  &  2.5 & 0.0068 & 0.26 & 0.0027 & 0.10 & 59 \\
10 & 82   & 33  & 0.1 & 9.1 & 7.4  & 18.5  &  --  &  2.5 & 0.009  & 0.34 & 0.0036 & 0.14 & 59 \\
10 & 164  & 66  & 0.3 & 9.1 & 3.7  & 9.3   &  --  &  2.5 & 0.018  & 0.68 & 0.0072 & 0.27 & 59 \\
10 & 328  & 131 & 0.5 & 9.1 & 1.9  & 4.6   & 5.2  &  2.5 & 0.036  & 1.4  & 0.014  & 0.55 & 59 \\
15 & 62   & 25  & 0.3 & 20.5 & 6.5  & 16.5  & 16.0 &  2.5 & 0.01   & 0.39 & 0.0041 & 0.15 & 59 \\
15 & 82   & 33  & 0.5 & 20.5 & 4.9  & 12.4  & 13.0 &  2.5 & 0.014  & 0.51 & 0.0054 & 0.20 & 59 \\
15 & 164  & 66  & 0.9 & 20.5 & 2.5  & 6.2   & 6.5  &  2.5 & 0.027  & 1.0  & 0.011  & 0.41 & 59 \\
15 & 328  & 131 & 1.9 & 20.5 & 1.2  & 3.1   & 3.3  &  2.5 & 0.054  &  2.0 & 0.022  & 0.82 & 59 \\
20 & 62   & 25  & 0.8 & 36.4 & 4.9  & 12.4  & 12.4 &  2.5 & 0.014  & 0.52 & 0.0054 & 0.20 & 59 \\
20 & 82   & 33  & 1.1 & 36.4 & 3.7  & 9.3   & 9.5  &  2.5 & 0.018  & 0.68 & 0.0072 & 0.27 & 59 \\
20 & 164  & 66  & 2.2 & 36.4 & 1.9  & 4.6   & 4.8  &  2.5 & 0.036  & 1.4  & 0.014  & 0.55 & 59 \\
20 & 328  & 131 & 4.4 & 36.4 & 0.9  & 2.3   & 2.5  &  2.5 & 0.072  & 2.7  & 0.029  & 1.09 & 59 \\
20 & 62   & 25  & 0.8 & 36.4 & 4.9  & 12.4  & 13.0 &  1.5 & 0.014  & 0.3  & 0.0054 & 0.12 & 42 \\
20 & 62   & 25  & 0.8 & 36.4 & 4.9  & 12.4  & 12.3 &  4.1 & 0.014  & 0.84 & 0.0054 & 0.33 & 89 \\
\hline
\end{tabular}
\end{table}

\begin{figure}[ht]
\includegraphics[width=0.49\textwidth]{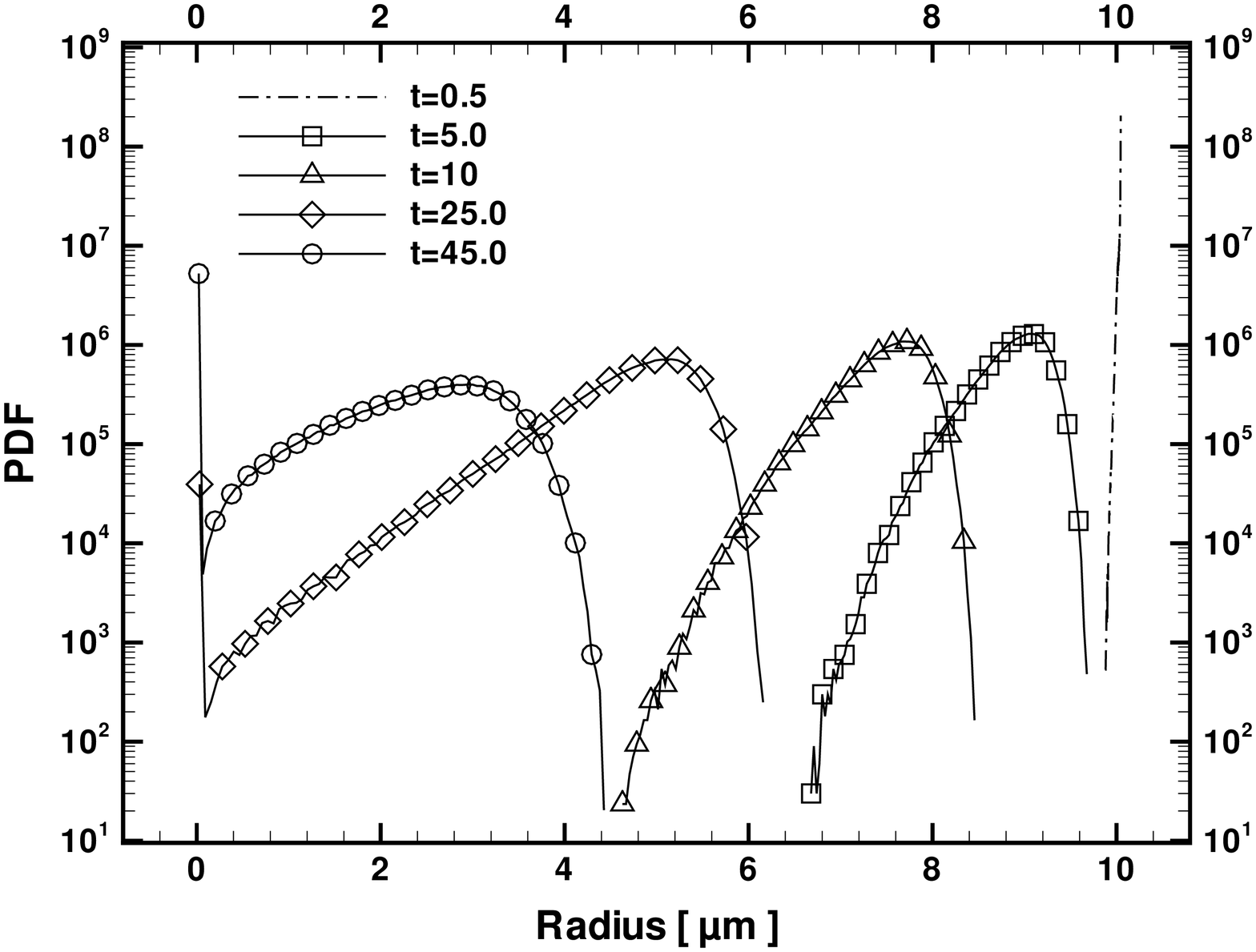}
\hspace{5mm}
\includegraphics[width=0.49\textwidth]{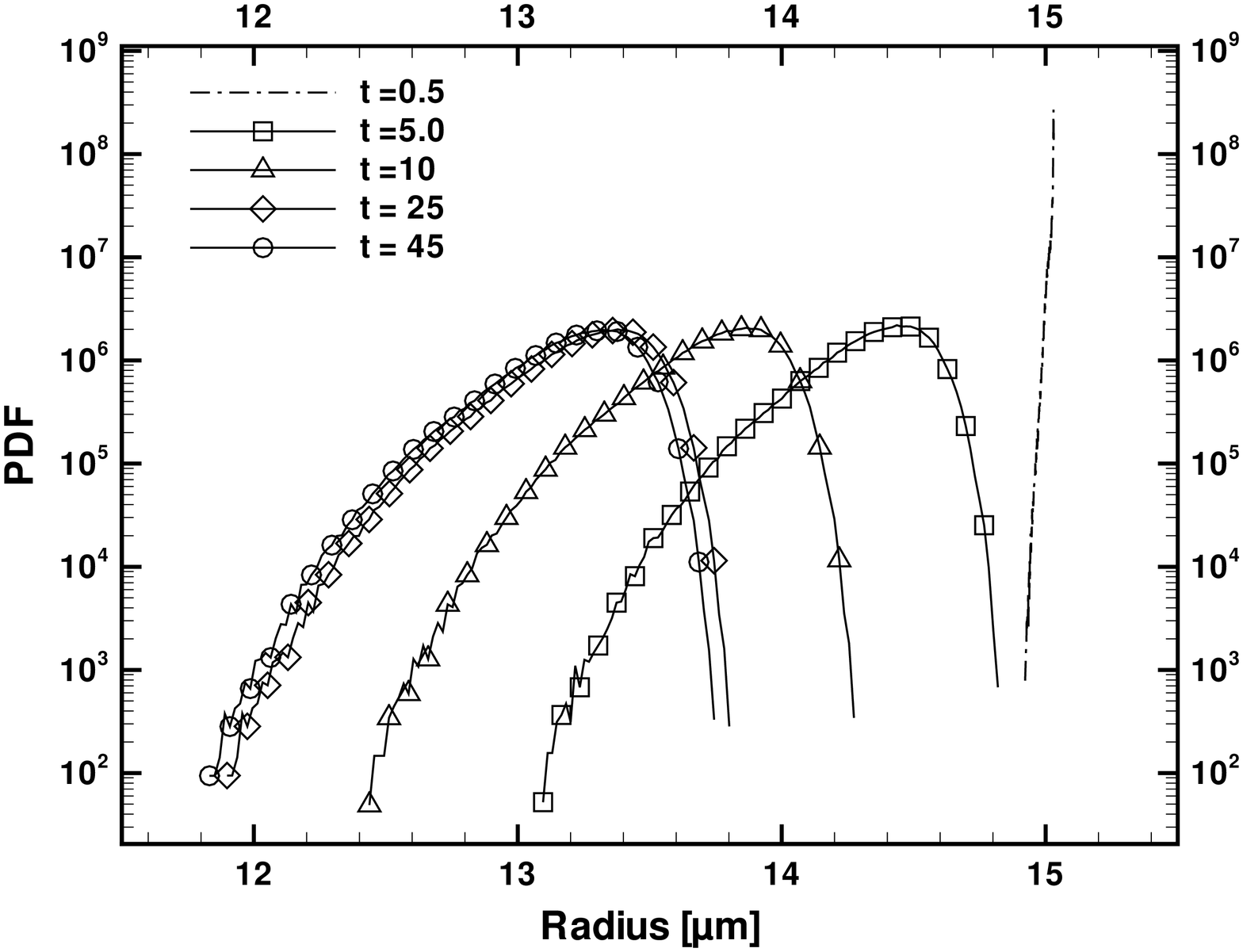}
\caption{Evolution of the droplet radii. The left panel represents evolution of the PDF for an
initial radius  $R_0 = 10 \mu m$ (Case 1) and right panel for  $R_0 = 15 \mu m$ (Case 2).
Both simulations started with total number of droplets of $N=1.100.000$. Data are for
$N_x^3=256^3$ and $R_{\lambda}=59$. }
\label{fig:R_PDF}
\end{figure}

Figure \ref{fig:R_PDF} shows the evolution of the droplet size distribution for examples of the two scenarios just described: $R_0 = 10 \mu m$ and $R_0 =15 \mu m $ with droplet number density $n_d = 164
cm^{-3}$. The left panel of the figure illustrates the evolution of the size distribution for $R_0 =
10 \mu m $ and indicates that after $t=25 s$  the first droplets have been evaporated. At this point the size distribution has developed a pronounced negative tail that appears nearly exponential.

After a minute, the majority of the droplets have evaporated. This continues until the last
droplet is vanished (not shown), corresponding to Case 1.  The right panel of the same figure illustrates the evolution of the size distribution for $R_0 = 15 \mu m$, corresponding to Case 2.  After $60$ seconds, all the droplets have stopped shrinking or growing, leaving a relatively narrow, but negatively-skewed size distribution. Droplets will be further advected in the sustained turbulent flow, but they do not shrink or grow in the homogeneously mixed vapor field.  In both examples the negative skewness of the size distribution is primarily a result of the nonuniform exposure of droplets to subsaturated air.

\begin{figure}
\centering
 \includegraphics[width=0.7\textwidth]{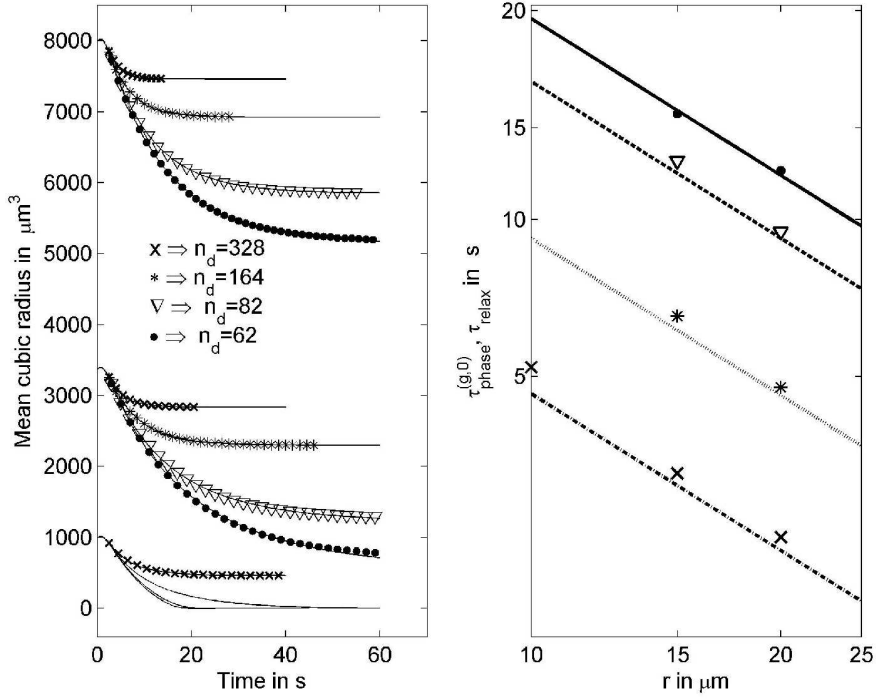}
\caption{The phase relaxation of the droplet ensemble. Left: Mean cubic radius versus time for
different initial radii and droplet number densities (solid lines). They are directly proportional to the
changing liquid water mass in the volume. Fits to the data in order to determine the corresponding
e-folding times are overlayed by symbols. Number densities are given in the legend. Right:
Comparison of the theoretical prediction for the phase relaxation time $\tau_{phase}^{(g,0)}$ (lines)
with the numerical value of $\tau_{relax}$ (symbols). Symbols agree with the left panel. All data are
$N_x^3=256^3$ and $R_{\lambda}=59$.}
 \label{fig:phase_time}
\end{figure}
\begin{figure}
\includegraphics[width=0.49\textwidth]{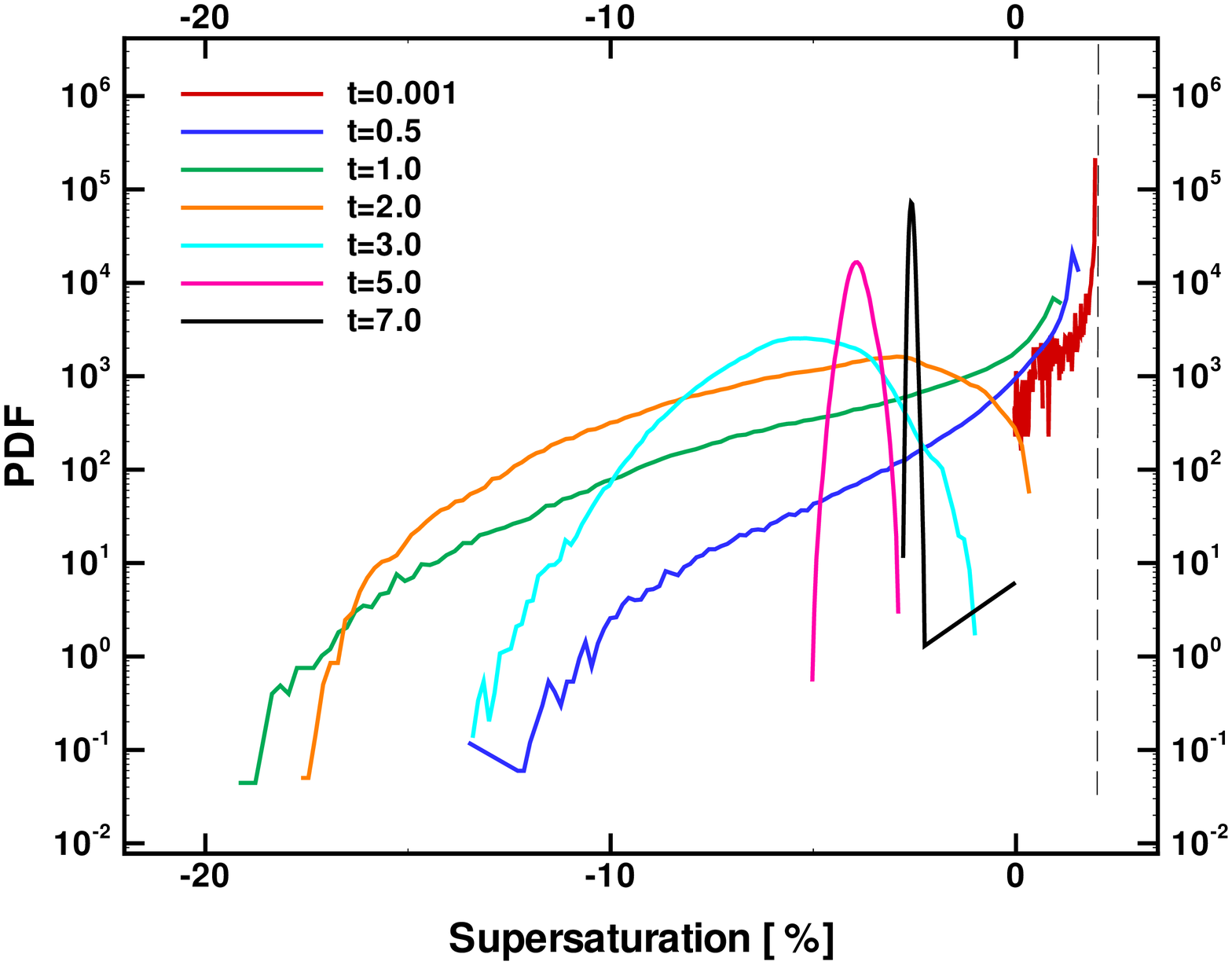}
\hspace{2mm}
\includegraphics[width=0.49\textwidth]{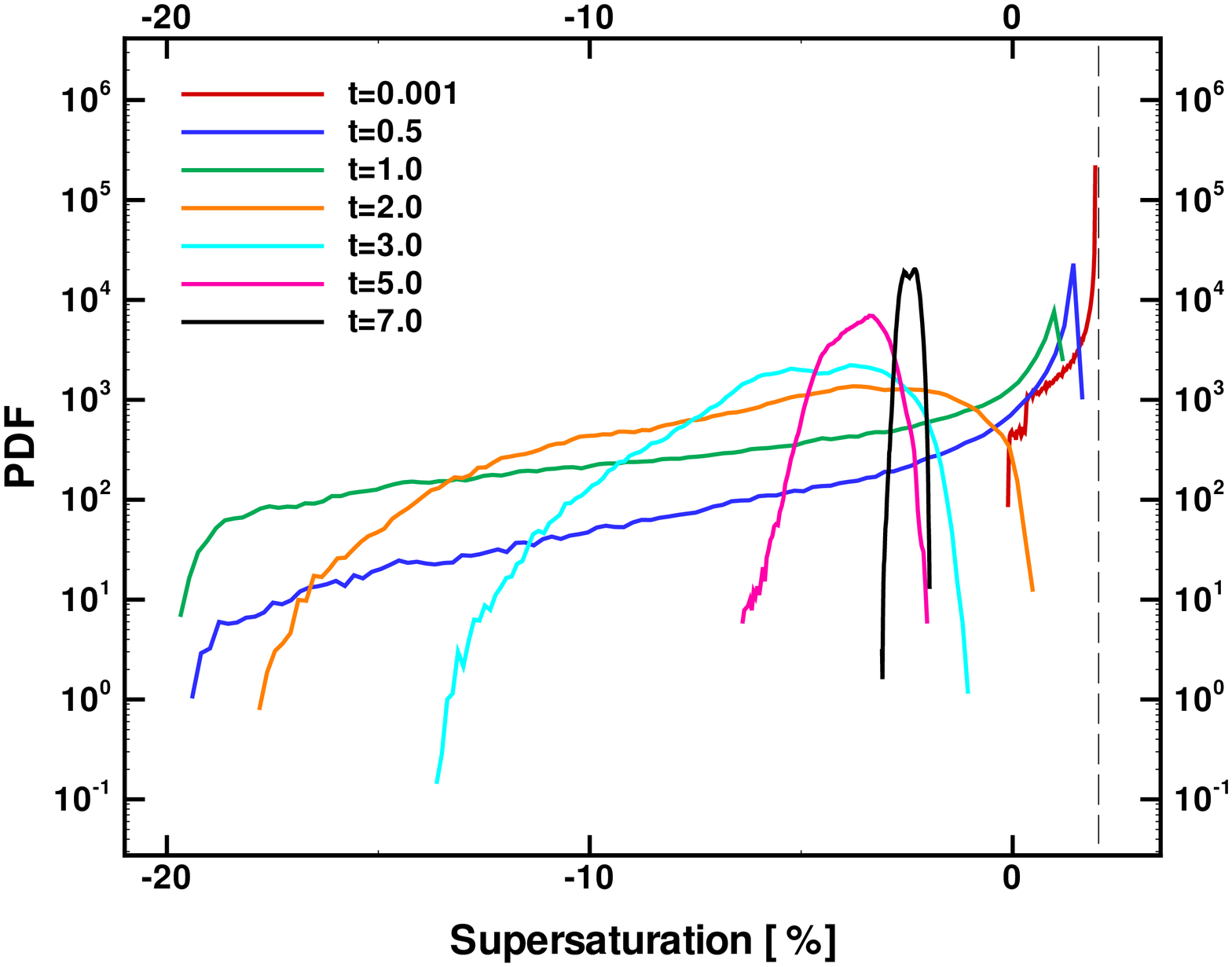}
\caption{Evaluation of the PDF of the supersaturation for the case of an initial radius of $R_0 = 20 \mu m$ and a droplet number density of $n_d = 164 cm^{-3}$. Data are again for $N_x^3=256^3$. At the beginning, supersaturation $S$ is 2\% and after 7 seconds it relaxes to about 2.5\% subsaturation. Left: without particle inertia and gravitational settling. Right: with particle inertia and gravitational settling.}
\label{fig:SuperS_PDF}
\end{figure}

Time scales have been of central interest in the discussion of inhomogeneous versus homogeneous mixing
for decades \cite{Baker1984,Jensen1989,Andrejczuk2004,Lehmann2009}, and there has been some
disagreement about which time scales correctly represent the microphysical response (e.g., see \cite{Andrejczuk2009}). We address the question here directly. The steady-state saturation level in Case 2 is reached with an approximately exponential relaxation time $\tau_{relax}$, which we compare with the phase relaxation time $\tau_{phase}$ as given by Eq. (\ref{tauphase}) and the single-droplet evaporation time $\tau_r = -R_0^2/2KS_0$ (the latter obtained via integration of Eq. (\ref{flux8}) assuming constant $S=S_0$). For times larger than $\tau_{relax}$ the cubic mean droplet radius $\langle r^3\rangle_L$, which is directly proportional to $M_l(t)$, becomes constant as illustrated in Fig. \ref{fig:phase_time} (left). The symbol $\langle\cdot\rangle_L$ denotes an average over the Lagrangian droplet ensemble. The relaxation time of the droplet ensemble $\tau_{relax}$ is obtained by a fit of a decaying exponential to the simulation graphs of the mean cubic radius. The fits to the data are indicated by the symbols in the left panel. The various time scales for all simulations are given in Tab. \ref{tab:2}. It is immediately evident that neither the single droplet evaporation time $\tau_r$ nor the cloud phase relaxation time based on the cloud values $\tau^{(c,0)}_{phase}$ correctly accounts for the observed relaxation. Figure \ref{fig:phase_time} (right) displays, instead, the comparison of the findings for $\tau_{relax}$ with the phase relaxation time $\tau^{(g,0)}_{phase}$ as  calculated from (\ref{tauphase}) but using the global number density $n_d^{(g,0)}$. As shown in Fig. \ref{fig:phase_time} (right), the indirect proportionality $\tau_{relax}\sim r^{-1}$ as given by the theoretical
prediction is confirmed. The same holds for $\tau_{relax}\sim n_d^{-1}$. One can observe that the relaxation times $\tau_{relax}$  are slightly greater than the corresponding values of $\tau^{(g,0)}_{phase}$ in most runs, presumably due to the steadily decreasing radius and the finite rate at which the droplets are spread throughout the volume.  It should be noted, however, that calculating a $\tau_{phase}$ based on the global number density and the \textit{final} droplet radius was less consistent with the observed $\tau_{relax}$.  The simple phase relaxation model is therefore a surprisingly good representation of the microphysical response to the mixing process in the range of Damk\"ohler numbers investigated. This range just barely approaches $Da_L \approx 1$,
and so is most representative of homogeneous mixing, just approaching the transition stage in the simulation with the highest liquid water content.

In order to understand the small quantitative disagreement, one has to recapitulate the assumptions
that enter the derivation of the phase relaxation time in (\ref{tauphase}). There, the droplet is embedded in a homogeneous vapor field, an assumption that is not fully sustained in the entrainment simulation. The different Lagrangian history of each individual droplet and the permanently changing saturation conditions cause a slower relaxation than the idealized situation assumed in the derivation of the phase relaxation time. As mentioned  before,  the initial vapor profile has a maximum amplitude of the supersaturation of $2\%$. Due to entrainment process, the value of supersaturation  starts decreasing from the beginning.  Figure \ref{fig:SuperS_PDF} (left panel) depicts the PDF of the supersaturation $S({\bf X},t)$ along the Lagrangian cloud droplet paths, monitored at different times up to $7$ seconds. The figure indicates that in the first seconds of the evolution the left tail of the PDF steadily grows, becoming approximately exponential. This initial time corresponds with the time required for clear air to reach the center of the original slab cloud, i.e., the large eddy time $T$.  It is consistent with the view given by Fig. \ref{fig:4}, in which a subset of droplets are mixed into the clear air and therefore experience stronger evaporation than the average.  Afterwards the left tail of the PDF narrows continuously until all droplets have reached the same subsaturation level of about $-2.5\%$ in this particular example.  Ultimately, the transient, nonuniform exposure during the early mixing leads to the negatively skewed size distributions shown in Fig. \ref{fig:R_PDF}, still preserved long after the transients have decayed.

\section{Reynolds number dependence of the entrainment}

The Reynolds number dependence of the relaxation process is studied in the following way: we prepared initial profiles for $n_d=62 cm^{-3}$ and $R_0=20\mu m$ in volumes of side lengths $L_x=
12.8,\,25.6$ and 51.2 $cm$. The initial vapor profile (\ref{profile1}) obeys the same parameters in all
three runs. Thus the volume of the initial slab cloud and the droplet number $N$ increase by a factor
of 8 and 64 when going from $L_x=12.8\,cm$ to $L_x=25.6\,cm$ and $L_x=51.2\,cm$, respectively.

\begin{figure}[h]
\centering
\includegraphics[width=0.6\textwidth]{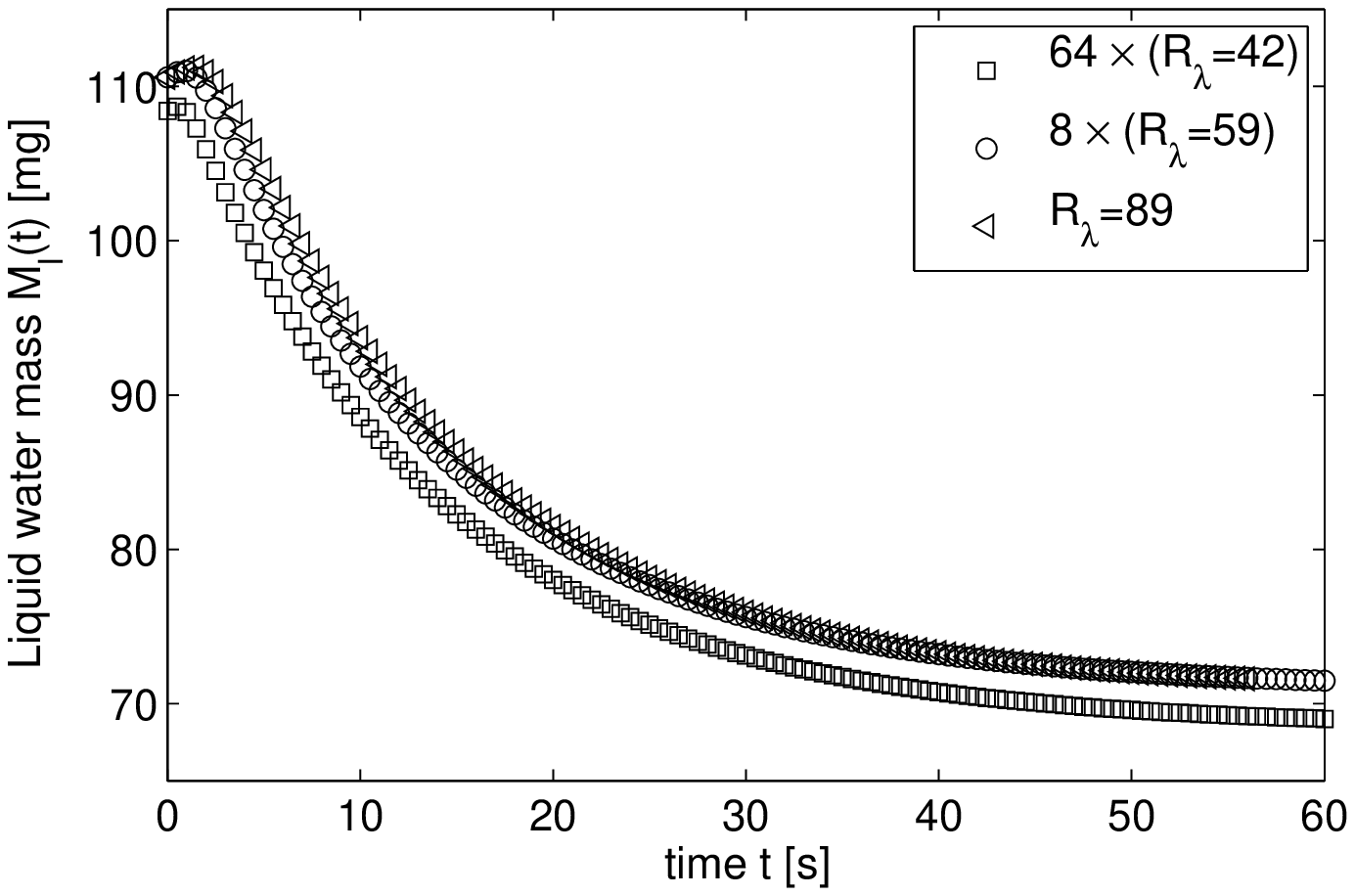}
\caption{Decay of the total water mass $M_l(t)$ versus time. The curves of the two smaller Reynolds
numbers have been rescaled by a factor of 8 and 64 in order to collapse the curves.}
\label{watermass}
\end{figure}

Figure \ref{watermass} indicates that the decay and thus the relaxation time differ only slightly with
increasing Reynolds number. This holds particularly for the final phase of the relaxation. Slight differences can be observed in the initial phase of the entrainment. The growth of the droplets is most
pronounced for the simulation in the biggest domain. The reason is that the ratio of entrainment area to volume is the smallest. Thus more droplets can grow unperturbed. As in the last section, the fit of
an exponential profile to the graphs accounts for this difference and includes the part of the curves only in which $M_l(t)$ decays monotonically. The relaxation times obtained are $\tau_{relax} = 12.3 s$
for $R_{\lambda}=89$,  $\tau_{relax} = 12.4 s$ for $R_{\lambda}=59$ and $\tau_{relax} = 13.0 s$ for $R_{\lambda}=42$, as given in Tab. \ref{tab:2}. The relaxation becomes slightly faster as the Reynolds number is increased.  By comparison, the global phase relaxation time for all three simulations is $\tau^{(g,0)}_{phase} = 12.4 s$.

\section{Role of particle inertia and gravitational settling}
The results of the droplet evolution presented in the previous section were obtained from a
Lagrangian model without particle inertia and gravitational settling. Thus, the particle velocities are exactly the same as the velocities of the surrounding fluid flow, i.e. ${\bf V}(t)={\bf u}({\bf X},t)$.
In the next step, all simulations have been repeated with gravitational settling and a finite particle response to variations of the local advecting velocity. This results in solving the full set of Eqns. (\ref{eqn:Lag_position})--(\ref{eqn:Lag_radius}). The maximum Stokes numbers which are obtained in the simulations are of the order of $St_{\eta}\lesssim 9\times 10^{-2}$ for $R_0=20\mu m$, and the corresponding settling parameter is $Sv_{\eta} = St_{\eta} (g/a_{\eta}) \lesssim 4$, where $a_{\eta}$ is the Kolmogorov acceleration \cite{Siebert2010}. Therefore, we expect that effects of particle inertia and gravitational sedimentation are just becoming significant for the largest droplet sizes considered.
\begin{figure}[h]
\centering
\includegraphics[width=0.49\textwidth]{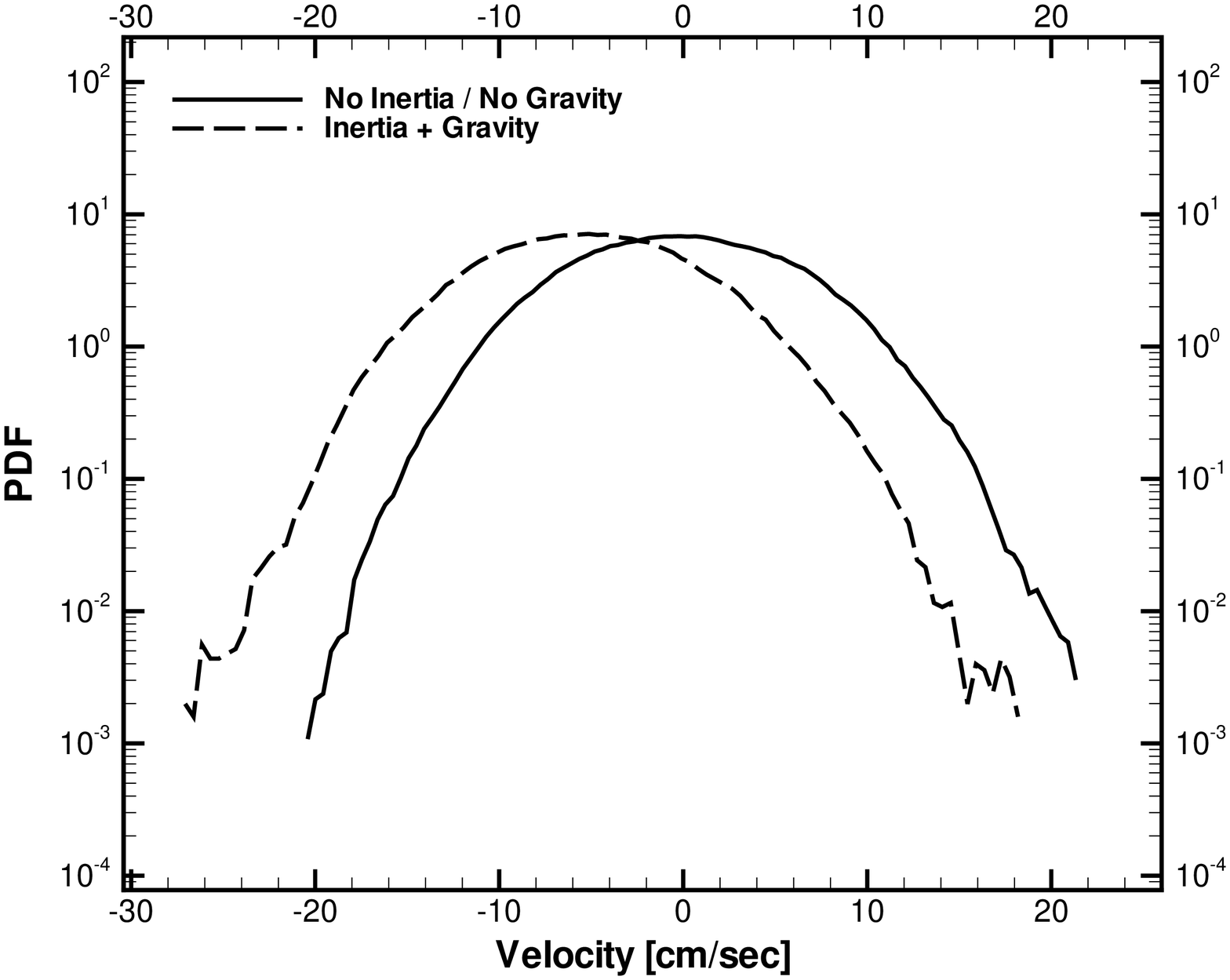}
\includegraphics[width=0.49\textwidth]{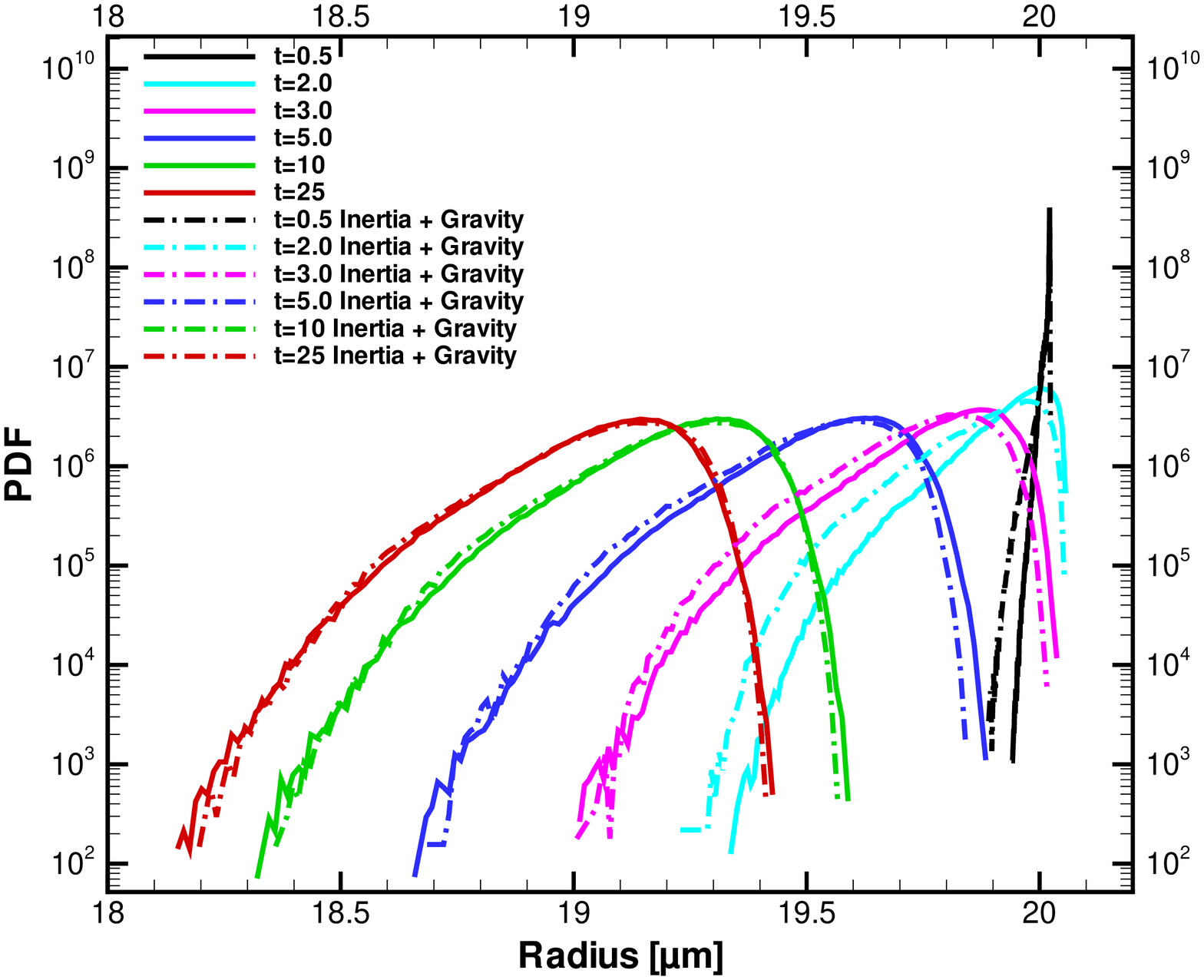}
\caption{Effects of particle inertia. Left: The PDF of the vertical Lagrangian velocity
component at $t = 10 s$  with and without  inertia effect. Right: Droplet
size distributions taken at four different times. Solid line graphs represent size distributions
with inertia and graphs with symbols represent the corresponding data without particle
inertia as indicated in the legend. In both panels we took $R_0=20 \mu m$ and $n_d=164
cm^{-3}$. All data are again for $N_x^3=256^3$ and times in seconds.}
\label{fig:R20_Inr_no_Inr_Paper}
\end{figure}

The relatively simple dynamical change due to gravity is illustrated in the PDF of the vertical droplet velocity component, shown in the left panel of Fig. \ref{fig:R20_Inr_no_Inr_Paper}. Two
distributions with and without inertia are depicted for $R_0=20\mu m$ and $n_d=164 cm^{-3}$.
There is an offset to a small negative amplitude that stands for the slow downward motion of the droplets, but the shape of the PDF is essentially unchanged. From the right panel, we can conclude
that the mean radius of the droplets relaxes to a value of about $\langle r\rangle_{\infty}\approx
19\mu m$ which corresponds with an inertial time scale of $\tau_p\approx 0.005 s$; the terminal
velocity estimate is $v_g=-g \tau_p \approx -5\,cm\,s^{-1}$, which is consistent with the offset
observed in the right panel of the figure.

But the microphysical effects of inertia and gravitational settling are more subtle.  As shown in the right panel of Fig. \ref{fig:SuperS_PDF}, the initial evolution of the supersaturation PDF is significantly altered by the presence of droplet inertia and gravity: the left tail of the PDF grows more rapidly, presumably as a result of stronger droplet decoupling from the fluid, but ultimately the supersaturation PDF collapses into a similar, relatively narrow but symmetric distribution. Droplets are initially exposed to very different vapor environments as a result of their inertia and gravitational settling, but ultimately the primary influence is on the evolution of the negatively skewed tail of the supersaturation pdf, and the overall homogenization of the supersaturation field is unchanged. A comparison of the droplet size distributions further unravels the systematic difference between the droplet dynamics with and without both effects.  Figure \ref{fig:R20_Inr_no_Inr_Paper} (right) indicates a more rapid evaporation and thus a more rapid pronounced left tail of the droplet size distribution. At later times, both become very similar but with slight depletion in both tails.
This means that the droplet evaporation is enhanced initially  when gravitational settling and inertial effects are present,  presumably because these effects lead to more rapid decoupling of droplets from the fluid containing high supersaturation.  We verified, however, that the variation of the mean radius with respect to time is almost unchanged when inertia is included.  This also can be interpreted as a result of the more rapid decorrelation of particles from the fluid, such that droplets that do come into contact with anomalously low supersaturations, do so for shorter times.  The present Lagrangian approach is particularly well suited to unraveling these details of the entrainment process. To summarize this part, it is found that particle inertia and gravitational settling have a strong influence on the initial evolution of the supersaturation PDF and on the positive and negative tails of the droplet size distribution, but that the transient effect is sufficiently small as to have essentially no influence on the mean droplet size.

\section{Summary and outlook}
A model for turbulent mixing and entrainment, that couples the Eulerian description of the velocity
${\bf u}$ and water vapor mixing ratio $q_v$ with a Lagrangian ensemble of cloud water droplets
has been presented in this paper with an emphasis on understanding the dynamics of the turbulent entrainment at the  interface between clear and cloudy air.  The direct numerical simulation model, which resolves the turbulence in a small subvolume of the cloud down to the Kolmogorov length $\eta$, is capable of generating
turbulent flow conditions as observed in cumulus clouds \cite{Lehmann2009}, e.g., low mean values of the kinetic energy dissipation rate are consistently obtained. Microphysical and turbulence parameters have been chosen to explore
the two limiting cases of turbulent mixing in this setup, homogenous and inhomogeneous mixing.
A central quantity to describe this physical process is the phase relaxation time, which has to be compared with the continuum of turbulence time scales.

Two basic dynamical scenarios are possible in the present setup, depending on the amount of
liquid water present at the beginning of the entrainment process. The first case leads to a complete evaporation of the droplets, and the second ends with a steady state droplet population surrounded by a saturated homogeneous vapor field. The relaxation time $\tau_{relax}$ to this state (which is obtained from the simulations) is compared
with the phase relaxation time $\tau_{phase}$. The observed $\tau_{relax}$ display the expected dependencies on $n_d$ and $r$
from Eq. (\ref{tauphase}).  The magnitude of $\tau_{relax}$, however, is found to be significantly larger
than $\tau_{phase}$ in all simulations, where the phase relaxation time is calculated with the undiluted cloud droplet number density (as is customary in the literature, to our knowledge). In contrast, very close agreement with the observed $\tau_{relax}$ is obtained when $\tau_{phase}$ is calculated using the diluted (or `global') number density.

The Lagrangian approach has allowed for detailed analysis of the droplet size distribution in conjunction with the evolution of the supersaturation field sampled at the droplet locations (supersaturation PDF). During the transient mixing event the initially perfectly monodisperse droplet population broadens significantly, with a distinct negative skewness.  This is partially a result of the strongly negatively skewed supersaturation PDF, which at early times in the mixing displays nearly exponential tails on the negative side of the distribution. This skewness arises from the droplets at the interface of the cloud, that are suddenly mixed into the clear air. Interpreted another way, this is an early manifestation of microphysical effects of inhomogeneous mixing, in which a subset of droplets is assumed to evaporate completely, leaving the remainder of droplets unchanged. These simulations have been performed primarily with $Da_L \lesssim 1$, i.e., favoring homogeneous mixing conditions, but not in the strongly homogeneous $Da_L \ll 1$ limit.

Effects of particle inertia and gravitational settling on the droplet size distribution and vertical particle velocities have also been analysed in the DNS, with Stokes numbers not exceeding $St_{\eta} < 9\times 10^{-2}$.  The primary influence is on the initial evaluation of the negative tail of the Lagrangian supersaturation PDF and the resulting acceleration  of the droplet evaporation. 
Within the parameter range studied, the mean droplet size was not modified by droplet inertia and settling. Likewise, mean droplet properties were not significantly altered with modest increases in Reynolds number.

In this work, the whole study has been conducted in a small subvolume of clear air-cloud interface, specifically in a cubic box of dimensions up to $L_x=51.2 \ cm$. For such size, we expect the mixing
to be dominantly homogeneous, with large-eddy Damkoehler numbers up to $Da_L$=2.7. This is exacerbated by the finding that the relevant phase relaxation time depends on the diluted droplet number density, so that the largest $Da_L$ only reach 1.1. In the future, we intend to carry out a similar analysis in a larger box such that inhomogeneous mixing can take over at the larger scales of the flow. This will allow us also to make contact with recent large-eddy simulations \cite{Steinfeld2008}.

Furthermore, it can be expected that the inclusion of the active character of the temperature field will modify the droplet growth. In the present case the temperature was set constant thereby reducing to an advection-diffusion problem for the vapor field. Full thermodynamic consistency  will require advection of temperature in the same turbulence, full consideration of latent heating associated with phase changes, and determination of the saturation vapor mixing ratio ratio as a function of the varying temperature field. Results of the present studies must therefore be interpreted in light of these simplifications, and considered to be a first step in building up to the full complexity of the cloud mixing problem.

\acknowledgements
We thank H. Siebert for helpful discussions on the initial conditions for the simulations. The authors acknowledge support by the Deutsche Forschungsgemeinschaft (DFG) within the Research Focus Program Metstr\"om (SPP 1276). JS  acknowledges additional support by the Heisenberg Program under Grant No. SCHU 1410/5-1. Furthermore, support from COST Action MP0806 is kindly acknowledged. The numerical simulations have been carried out at the J\"ulich Supercomputing Centre (Germany) under Grant No. HIL03. RAS acknowledges support from National Science Foundation Grant No. AGS1026123.  RAS and JS  were also supported in part by the National Science Foundation under Grant No. PHY05-51164 within the program ``The Nature of Turbulence'', held at the Kavli Institute of Theoretical Physics at the University of California in Santa Barbara.




\end{document}